\documentclass[11pt,english,openbib,leqno,epsfig]{article}
\usepackage[T1]{fontenc}
\usepackage[latin1]{inputenc}
\usepackage[letterpaper]{geometry}
\usepackage{amsmath}
\usepackage{graphicx}
\usepackage{amssymb}
\usepackage{esint}



\usepackage{babel}

\makeatother

\usepackage{babel}

\makeatother

\usepackage{babel}

\makeatother

\usepackage{babel}

\makeatother

\usepackage{babel}

\makeatother

\usepackage{babel}

\begin{document}

\title{Invariant measures of the 2D Euler and Vlasov equations}

\author{Freddy Bouchet$^{1,2,3,}$\footnote{Author to whom any correspondence should be addressed.} \ and Marianne Corvellec$^{1}$}

\maketitle
Affiliations:

$^{1}:$ INLN, CNRS, UNS, 1361 route des Lucioles, 06560 Valbonne, France

$^{2}:$ CNLS, LANL, MS B258 Los Alamos National Laboratory, PO Box 1663,
Los Alamos, NM 87545, United States

$^{3}:$ Laboratoire de Physique, ENS de Lyon, CNRS, Universit\'e de Lyon,
46 all\'ee d'Italie, 69364 Lyon cedex 07, France 
\par\smallskip
Email: $^{*}$Freddy.Bouchet@ens-lyon.fr, Marianne.Corvellec@ens-lyon.fr
\par\bigskip
%Received 22 March 2010 at J. Stat. Mech.\\
%Accepted 11 July 2010\\
%Published 26 August 2010 as J. Stat. Mech. (2010) P08021\\
%Online at http://iopscience.iop.org/1742-5468/2010/08/P08021

%doi: 10.1088/1742-5468/2010/08/P08021

\begin{abstract}
We discuss invariant measures of partial differential equations such
as the 2D Euler or Vlasov equations. For the 2D Euler equations, starting
from the Liouville theorem, valid for $N$-dimensional approximations
of the dynamics, we define the microcanonical measure as a limit measure
where $N$ goes to infinity. When only the energy and enstrophy invariants
are taken into account, we give an explicit computation to prove the
following result: the microcanonical measure is actually
a Young measure corresponding to the maximization of a mean-field
entropy. We explain why this result remains true for more general
microcanonical measures, when all the dynamical invariants are taken
into account. We give an explicit proof that these microcanonical
measures are invariant measures for the dynamics of the 2D Euler equations.
We describe a more general set of invariant measures, and discuss
briefly their stability and their consequence for the ergodicity of
the 2D Euler equations. The extension of these results to the Vlasov
equations is also discussed, together with a proof of the uniqueness
of statistical equilibria, for Vlasov equations with repulsive convex
potentials.

Even if we consider, in this paper, invariant measures only for Hamiltonian
equations, with no fluxes of conserved quantities, we think this work
is an important step towards the description of non-equilibrium invariant
measures with fluxes. 
\end{abstract}

\section{Introduction}

In a complex and chaotic dynamical system, such as a Hamiltonian
system with a large number of degrees of freedom, or a non-equilibrium
steady state where dissipation balances forcing on average, the knowledge
of a non-trivial invariant measure is equivalent to the knowledge
of the statistical properties of all physical quantities. It is thus an essential concept. In a turbulent problem, the knowledge
of an invariant measure gives access to the stationary probability
distribution function of all physical quantities, and gives a solution
to the usual hierarchy of the $n$-point correlator of the velocity
field, among other things. A series of very interesting recent works
have proved the existence of invariant measures and described some
of their properties, for instance in stochastic systems forced by
noises \cite{Kuksin_2004_JStatPhys_EulerianLimit,Kuksin_Shirikyan_2000_CMaPh,Mattingly_Sinai_1999math_3042M,Bricmont_Kupianen_2001_Comm_Math_Phys_Ergodicity2DNavierStokes}.
However, only in very few instances of complex systems, is an invariant
measure explicitly known. Finite-dimensional Hamiltonian systems are
among these exceptions: thanks to the Liouville theorem, a uniform
measure on a constant-energy shell of phase space is invariant (microcanonical
measure); the canonical Gibbs measures are other explicit examples
of invariant measures. This essential remark is at the base of equilibrium
statistical mechanics.

For Hamiltonian partial differential equations, the situation is more
complex. Indeed, the dimension of the system is then infinite. Thus,
the meaning of phase space volume and the proper normalization of
a uniform measure over a constant-energy shell of phase space are
not clear notions. Microcanonical or Gibbs measures then have to be
built carefully and their properties have to be checked. There are
a few examples, where a Gibbs-type invariant measure has been proved
to exist (see for instance \cite{Bourgain_1994_CMP,Bourgain_2000_CMP}
in the case of the nonlinear Schrödinger equation, see also
\cite{Lebowitz_Rose_Speer_1988JSP,McKean_1995CMaPh}).
As in \cite{Bourgain_1994_CMP,Bourgain_2000_CMP}, such a proof usually
involves the study of approximations of the partial differential equations
with finite dimension $N$, and of limits of ensembles of measures
when $N$ goes to infinity. That the limit is actually an invariant
measure of the initial partial differential equation completes the
proof. We consider in this paper the construction of microcanonical
and Gibbs measures for the 2D Euler equations and for the Vlasov equations.
We also consider other sets of invariant measures for these equations,
and investigate their dynamical stability.\\

The flow of a perfect fluid is described by the Euler equations, one
of the oldest equations in mathematical physics \cite{euler250}.
More than two and a half centuries after their discovery by Euler, these equations
still offer great challenges to both mathematicians and physicists
\cite{euler250}. Two-dimensional flows and the two-dimensional Euler
equations are mathematically much simpler than their three-dimensional
counterparts, but still present some very interesting unsolved problems.
The strong analogies between the 2D Euler equations and the Vlasov
equations have been observed at least since the `50s: they are both
nonlinear transport equations, the non-linearity being due to non-local
and non-integrable (at long distance) interactions (long-range interactions) %
\footnote{Whereas in the 2D Euler and Vlasov equations, the long-range interactions
and the associated non-additivity are essential for thermodynamical
properties of the system (leading possibly to statistical ensemble
inequivalence), they are not essential in the following discussion
dealing with invariant measures and dynamics. For instance all the
discussion of this paper can be easily adapted to the Quasi-Geostrophic
model for which the interaction decays exponentially for distance
much larger than the Rossby deformation radius.%
}. We hope that this work may be very useful in the large number of
systems with long range interactions that have been studied recently
\cite{Bouchet_Barre:2005_JSP,Dauxois_Ruffo_Arimondo_Wilkens_2002LNP...602....1D,Campa_Dauxois_Ruffo_Revues_2009_PhR...480...57C,Bouchet_Gupta_Mukamel_PRL_2009,Chavanis_2006IJMPB_Revue_Auto_Gravitant}.

One of the main physical phenomena arising from these two equations
is the self-organization into large-scale coherent structures: large-scale
particle clusters for the Vlasov equations with attractive potential,
large-scale particle clouds (whose density profiles depend on both
the dynamical invariants and the external confining potential) for
the Vlasov equations with repulsive potential, or monopoles, dipoles,
and parallel flows for the 2D Euler equations. Such large-scale structures
are analogous to geophysical cyclones, anticyclones, and jets in the
oceans and atmospheres \cite{Bouchet_Sommeria:2002_JFM}. This analogy,
understood thanks to the strong theoretical similarities between the
2D Euler equations on one hand, and the quasi-geostrophic or shallow-water
models on the other hand, is one of the main motivations for the study
of the 2D Euler equations. The 2D Euler equations also describe experimental
flows, such as the transverse dynamics of electron plasma columns
\cite{Schecter_Dubin_etc_Vortex_Crystals_2DEuler1999PhFl}, the large--Reynolds-number
approximation of the dynamics of fluids when three-dimensional motion
is constrained by a strong transverse field (rotation, transverse
magnetic field in a liquid metal, etc. \cite{Sommeria_1986_JFM_2Dinverscascade_MHD}),
or the dynamics of fluids in very thin geometries \cite{Paret_Tabeling_1998_PhysFluids}.

As first guessed by Onsager \cite{Onsager:1949_Meca_Stat_Points_Vortex},
such a self-organization is naturally explained by equilibrium statistical
mechanics: the infinite number of degrees of freedom involved should
have a macroscopic behavior that corresponds to an average over a
microcanonical measure. Because of the long-range interactions between
vortices, such an equilibrium is not uniform and depends strongly
on the boundary conditions. This explains the formation of vortices
and jets. In his work \cite{Onsager:1949_Meca_Stat_Points_Vortex},
Onsager circumvented the difficulty of infinite dimensions discussed
above, by studying the point-vortex model, which is a finite-dimensional
Hamiltonian system, describing the dynamics of singular vortices.
This model is actually a special class of solutions to the 2D Euler
equations (please see \cite{Chavanis_houches_2002,Chavanis_Lemou_2005_PRE,Dubin_2003_Phys_Plasma_Collisional_Diffusion_Point_Vortex}
and references therein for recent developments of the kinetic theory
of point-vortices). The equilibrium statistical mechanics of the point-vortex
model has a long and very interesting history, with wonderful pieces
of mathematical achievements \cite{Onsager:1949_Meca_Stat_Points_Vortex,Joyce_Montgommery_1973,CagliotiLMP:1995_CMP_II(Inequivalence),Kiessling_Lebowitz_1997_PointVortex_Inequivalence_LMathPhys,Dubin_ONeil_1988_PhysRevLett_Kinetic_Point_Vortex,Chavanis_houches_2002,Eyink_Spohn_1993_JSP....70..833E,Eyink_Sreenivasan_2006_Rev_Modern_Physics}.

The study of the equilibrium statistical mechanics of the 2D Euler
equations or of the Vlasov equations with usual smooth initial conditions
implies considering the real infinite-dimension nature of the problem.
An equilibrium statistical mechanics was proposed for these equations,
based on the maximization of a Boltzmann--Gibbs entropy constrained
by the conservation of the invariants, by Lynden-Bell for the Vlasov
case \cite{LyndenBell:1968_MNRAS}, and by Robert, Sommeria, and Miller
for the 2D Euler case \cite{Robert:1990_CRAS,Miller:1990_PRL_Meca_Stat,SommeriaRobert:1991_JFM_meca_Stat}
(see also \cite{Chavanis_etal_APJ_1996} for a discussion on the analogies
between both systems). Such an approach
is basically a mean-field one: the use of the Boltzmann--Gibbs entropy for the entropy
(or mean-field entropy) can be justified
with the classical Boltzmann counting argument, which is meaningful only if
correlations between the vorticity values (or values of the one-particle
distribution in the Vlasov case) at different points can be neglected.

Although this mean-field approach was a phenomenological assumption
in the first papers \cite{LyndenBell:1968_MNRAS,Robert:1990_CRAS,Miller:1990_PRL_Meca_Stat,Robert:1991_JSP_Meca_Stat},
it can be proved to be exact in systems with long-range interactions.
For instance, physicists discussed the validity of the mean-field
approach for the point-vortex model in the `80s, arguing that a Kramer-Moyal
expansion when $N$ goes to infinity is self-consistent. During the
`90s, mathematicians and mathematical physicists proved that the mean-field
approach is correct for the point-vortex model, using different tools
\cite{CagliotiLMP:1995_CMP_II(Inequivalence),Eyink_Spohn_1993_JSP....70..833E,Kiessling_Lebowitz_1997_PointVortex_Inequivalence_LMathPhys}.
For the 2D Euler and Vlasov equations, following the work of Bourgain
for the nonlinear Schrödinger equation, such a proof\textbf{ }should\textbf{
}involve three steps: 
\begin{enumerate}
\item Proving Liouville theorems for finite-dimensional approximations of
the dynamical equations, and defining microcanonical measures for
those, 
\item Studying the infinite-$N$ limit for this set of measures and proving
that the limit is actually described by the maximization of a mean
field entropy, 
\item Proving that the limit measure is a dynamically invariant measure
of the 2D Euler (or Vlasov) equations. 
\end{enumerate}
For Galerkin truncations of the 2D Euler equations, e.g., using Fourier
mode decomposition, Point 1 above is a classical result \cite{Lee52};
actually, the Euler equations verify a detailed Liouville theorem
\cite{Kraichnan_Motgommery_1980_Reports_Progress_Physics}. More recently,
Robert \cite{Robert_2000_CommMathPhys-TruncationEuler} proved that
a much larger class of approximations of the 2D Euler equations, obtained
by $L_{2}$ projections (e.g., finite-element approximations), verify
a Liouville theorem. This last point is very important, because measures
based on spatial truncations are much more natural than those based
on Fourier mode decompositions for systems with long-range interactions.
This makes the proof of Point 2 much easier. Indeed, Michel and Robert
\cite{Michel_Robert_LargeDeviations1994CMaPh.159..195M} proved large-deviation
results for ensembles of Young measures where the mean-field entropy
appears as the opposite of the large-deviation rate function. A similar large-deviation
result, justifying also the maximization of a mean-field entropy,
was obtained from sets of measures based on spatial discretization
\cite{Boucher_Ellis_1999_AP,Boucher_Ellis_Turkington_2000_JSP}, but
without any reference to the dynamics. These two results are essential
steps, but give only a partial answer to Point 2, as the relation
to dynamics is missing, making any further step towards Point 3 more
difficult. As noted in \cite{Robert_2000_CommMathPhys-TruncationEuler},
another difficulty is that Fourier or $L_{2}$ projections conserve
only the quadratic invariants. Then, finite-dimensional measures that
are based on the Liouville theorem and that take into account further
invariants, which is the relevant procedure, are not invariant for
the finite-dimensional approximation of the dynamical system. This
makes Point 3 out of reach. For instance, the derivation of Point
3 for the nonlinear Schrödinger equation by Bourgain \cite{Bourgain_1994_CMP,Bourgain_2000_CMP}
relies deeply on the fact that this equation has only quadratic and
linear invariants, and that finite-dimensional approximate measures
are invariants for the finite-dimensional dynamical approximations.

From this discussion, we conclude that although the validity of the
mean-field approximation and the validity of the maximization of a mean-field
entropy for some sets of natural measures have been proved \cite{Michel_Robert_1994_JSP_GRS,Boucher_Ellis_1999_AP,Boucher_Ellis_Turkington_2000_JSP},
a clear proof of the relation of those microcanonical measures with
dynamics is still missing. Another important question is to know
whether all measures built from any reasonable finite-dimensional
truncation would lead to a unique limit measure in the limit of infinite
dimension.\\

As regards the relation between microcanonical (or canonical) measures
and dynamics, we argue in this paper that it is not necessary to follow
the classical program described above (points 1, 2, 3). More precisely, we argue that
there is no logical need for the finite-dimensional approximate measures
to be invariant measures of the finite-dimensional approximations
of the dynamical system. We can indeed rely on the large-deviation
results \cite{Michel_Robert_1994_JSP_GRS,Boucher_Ellis_1999_AP} and
verify only\emph{ a posteriori }that the limit measure is an invariant
measure of the 2D Euler equations. This is what we do in this paper,
by studying directly the evolution equation for measures of the 2D
Euler equations and studying ensembles of invariant measures. The
discussion relies on Young measures, i.e., product measures for which
the vorticity values at different points in space are independent
random variables. The importance of Young measures in the context
of equilibrium statistical mechanics was stressed in \cite{Michel_Robert_LargeDeviations1994CMaPh.159..195M,Robert_2000_CommMathPhys-TruncationEuler}.
One of the main motivations for the present work is to give a dynamical
meaning to Young measures, that we guess will be essential for further
developments.

We write the evolution equation for the characteristic functional
of the vorticity field, for the 2D Euler dynamics. Applying these
equations to Young measures, we can study the class of Young measures
that are invariant. We then note that microcanonical (or canonical)
measures, which are actually Young measures, are dynamically invariant.
The class of invariant Young measures is however larger; we explain
that this property is related to a breaking of ergodicity for the
2D Euler equations. The dynamical approach of this paper also allows
to discuss the dynamical stability of invariant measures, and particularly
of invariant Young measures. The possibility to discuss dynamics is
one of the main motivations for this work, as will be further discussed
in the conclusion.\\

In this paper, we also give a partial answer to the second question
above: would any truncation lead to the same limit measure? For this
we consider the case of the energy-enstrophy measure (which takes
into account only the energy and enstrophy as dynamical invariants).
With direct explicit computations of the expectations for the Fourier
$N$-dimensional approximate measures, we prove that the limit measure
is the same as the one obtained by maximizing the mean-field entropy
(obtained as a large-deviation functional of a Young measure or from
spatially discretized systems).

Energy-enstrophy measures were first investigated by Kraichnan \cite{Kraichnan:1975_statisical_dynamics_2D_flow},
in the canonical ensemble, and without considering the limit of an
infinite number of degrees of freedom. In this paper, for the first
time, we derive explicitly the microcanonical measure, which is the
relevant physical one, for it takes into account the constraints explicitly.
This derivation is based on expressing the constraint as a complex
integral of an auxiliary variable, and analyzing precisely the integral
asymptotics through the saddle-point approximation. This allows to
actually take the limit of an infinite number of degrees of freedom
with a fixed value of the invariants. This is a good opportunity to
discuss again the Kraichnan theory: although the microcanonical measure
leads to the same qualitative prediction as Kraichnan's, i.e., the
condensation of energy in the largest-scale mode, we prove that microcanonical
and canonical ensembles are not equivalent. More precisely, this is
a situation of partial ensemble equivalence (see \cite{EllisHavenTurkington:2000_Inequivalence}
for a definition).\\

In section \ref{sec:Equilibrium-statistical-mechanics}, we introduce
the 2D Euler equations and their invariants, we define the microcanonical
measure, and we give a pedagogic heuristic account of the meaning
of the mean-field approximation and of the resulting statistical equilibrium
distributions.

In section \ref{sec:Energy-Enstrophy-microcan-meas}, we perform the
computations for the energy-enstrophy measure, both directly from
its definition through Fourier modes and from the mean-field variational
problem. We show that the results are equivalent and discuss ensemble
inequivalence as well as the Kraichnan theory.

Section \ref{sec:Invariant-measures-2D-Euler} is devoted to the dynamics
of measures for the 2D Euler equations. We characterize the set of
invariant Young measures and their stability. In section \ref{sec:Invariant-measures-Vlasov},
we briefly explain the generalization of these results to the Vlasov
equations. We also prove the uniqueness of statistical equilibria
in the case of convex repulsive interaction potentials.

The relation of this work to non-equilibrium problems is discussed
in section \ref{sec:Conclusion-and-perspectives}.

\section{Equilibrium statistical mechanics of two-dimensional flows\label{sec:Equilibrium-statistical-mechanics}}

\subsection{2D Euler equations}

Defining the vorticity as $\omega=\left(\nabla\times\mathbf{v}\right)\!\cdot\mathbf{e}_{z}\,,$
the 2D Euler equations take the simple form of a conservation law
for the vorticity. They read \begin{equation}
\frac{\partial\omega}{\partial t}+\mathbf{v}\boldsymbol{\cdot\nabla}\omega=0\,\,;\quad\mathbf{v}=\mathbf{e}_{z}\times\boldsymbol{\nabla}\psi\,\,;\quad\omega=\Delta\psi,\label{eq:Euler_2D_Vorticity}\end{equation}
 where the\textbf{ }solenoidal (incompressible)\textbf{ }velocity
$\mathbf{v}$ is expressed as the orthogonal of the streamfunction
gradient $\nabla\psi$. We complement the equation $\omega=\Delta\psi$
with impenetrability boundary conditions: $\psi=0$ on $\partial\mathcal{D}$,
where $\mathcal{D}$ is a simply connected domain.

\smallskip{}
 The kinetic energy of the flow is conserved. It reads \begin{equation}
\mathcal{E}\left[\omega\right]=\frac{1}{2}\int_{\mathcal{D}}\mathrm{d}\mathbf{r}\,\mathbf{v}^{2}=\frac{1}{2}\int_{\mathcal{D}}\mathrm{d}\mathbf{r}\,\left(\nabla\psi\right)^{2}=-\frac{1}{2}\int_{\mathcal{D}}\mathrm{d}\mathbf{r}\,\omega\psi.\label{eq:Energy}\end{equation}

Other conservation laws are discussed in section \ref{sub:Casimirs-conservation-laws}.

\subsection{Microcanonical measures}

\subsubsection{Theoretical foundations of equilibrium statistical mechanics\label{sub:Hamiltonian-structure}}

Let us consider a canonical Hamiltonian system: $\{q_{i}\}_{1\leq i\leq N}$
denote the generalized coordinates, $\{p_{i}\}_{1\leq i\leq N}$ their
conjugate momenta, and $H(\{q_{i},p_{i}\})$ the Hamiltonian. The
variables \textbf{$\{q_{i},p_{i}\}_{1\leq i\leq N}$ }belong to a
$2N$-dimensional space $\Omega$ called the phase space. Each point
$(\{q_{i},p_{i}\})$ is called a microstate. The equilibrium statistical
mechanics of such a canonical Hamiltonian system is based on the Liouville
theorem, which states that the non-normalized measure \[
\mu=\prod_{i}\mathrm{d}p_{i}\mathrm{d}q_{i}\]
 is dynamically invariant. The invariance of $\mu$ is equivalent
to \begin{equation}
\sum_{i}\left(\frac{\partial\dot{q}_{i}}{\partial q_{i}}+\frac{\partial\dot{p}_{i}}{\partial p_{i}}\right)=0\,,\label{eq:Liouville}\end{equation}
 which is a direct consequence of the Hamilton equations of motion:
\[
\left\{ \begin{aligned}\dot{q_{i}} & =\frac{\partial H}{\partial p_{i}}\,,\\
\dot{p}_{i} & =-\frac{\partial H}{\partial q_{i}}\,.\end{aligned}
\right.\]
 Note that the equations of motion can also be written in a Poisson
bracket form: \begin{equation}
\left\{ \begin{aligned}\dot{q_{i}} & =\left\{ q_{i},H\right\} ,\\
\dot{p_{i}} & =\left\{ p_{i},H\right\} .\end{aligned}
\right.\label{eq:Crochets_Poisson_Canonique}\end{equation}
 Each term of the sum \eqref{eq:Liouville} actually vanishes independently:
\[
\forall\, i,\quad\frac{\partial\dot{q}_{i}}{\partial q_{i}}+\frac{\partial\dot{p}_{i}}{\partial p_{i}}=0\,.\]
Such a relation is called a detailed Liouville theorem.

\smallskip{}
For all conserved quantities $\left\{ I_{1}(q,p),\ldots,I_{n}(q,p)\right\} $
of the Hamiltonian dynamics, the measures \begin{equation}
\mu_{c}=\frac{1}{Z_{c}}\prod_{i}\mathrm{d}p_{i}\mathrm{d}q_{i}\, F\!\left(I_{1},\ldots,I_{n}\right)\label{eq:Canonical_measure}\end{equation}
are also invariant measures, $F$ being any function; $Z_{c}$ is
a normalization constant. An important question is to know which of
these measures are relevant for describing the statistics of the physical
system.

In the case of an isolated system, the dynamics is Hamiltonian and
there is no exchange of energy or other conserved quantities with
the environment. It is therefore natural to consider a measure that
takes into account all these dynamical invariants as constraints.
This justifies the definition of the microcanonical measure\textbf{
}(for a given set of invariants $\left\{ I_{1}(q,p),\ldots,I_{n}(q,p)\right\} $):
\begin{equation}
\mu_{m}\left(I_{1_{0}},\ldots,I_{n_{0}}\right)=\frac{1}{\Omega\left(I_{1_{0}},\ldots,I_{n_{0}}\right)}\prod_{i}\mathrm{d}p_{i}\mathrm{d}q_{i}\prod_{k=1}^{n}\delta\left(I_{k}(q,p)-I_{k_{0}}\right)\!,\label{eq:Microcanonical_Measure_H}\end{equation}
 where $\Omega(I_{1_{0}},\ldots,I_{n_{0}})$ is a normalization constant
--- for small $\left\{ \Delta I_{k}\right\} $, $\Omega(I_{1_{0}},\ldots,I_{n_{0}})\prod_{k=1}^{n}\Delta I_{k}$
is the volume of the part of phase space such that for all $k$, $I_{k_{0}}\leq I_{k}\leq I_{k_{0}}+\Delta I_{k}$
\footnote{A more natural definition of the microcanonical measure would be the
uniform measure on the submanifold defined by $I_{k}=I_{k_{0}}$ for
all $k$. This would require adding determinants in formula (\ref{eq:Microcanonical_Measure_H}),
and imply further technical difficulties. In most cases though, in
the limit of large number $N$ of degrees of freedom, these two definitions
of the microcanonical measure become equivalent. Indeed, the measures
have then large-deviation properties (saddle-point evaluations), where
$N$ is the large parameter, and such determinants become irrelevant.
We note that in the original works of Boltzmann and Gibbs,
the microcanonical measure refers to a measure where only the energy
constraint is considered.}%
. Then, the Boltzmann entropy of the Hamiltonian system is \[
S=k_{B}\log\Omega\,,\]

where $k_{B}$ is the well-known Boltzmann constant. When the system
under consideration is not isolated, but coupled with an external
bath of conserved quantities, other measures are necessary to describe
properly the system by equilibrium statistical mechanics. A classical
statistical mechanics result proves that the relevant functions $F$
in (\ref{eq:Canonical_measure}) are then exponentials (Boltzmann
factors), and measures of type (\ref{eq:Canonical_measure}) are referred
to as canonical measures (hence the `c' subscript). The notion
of a coupling with a single thermal bath (or to a bath of other conserved
quantity than energy) in statistical mechanics assumes that if the
system is coupled to a single bath, the total system (bath + system)
is described by a microcanonical measure. This is an essential assumption
in order to derive the canonical measure. By contrast, when systems
are coupled to the environment through irreversible couplings (without
detailed balance), or through several baths with different thermodynamic
parameters, there is no reason anymore to expect a canonical measure
to describe the statistics of the system. Fluid mechanics systems,
where usually the stirring processes and the dissipation mechanisms
are physical phenomena of a different nature, can never be considered
as coupled to a single bath of some conserved quantities (in all works
so far on fluid systems, where the distribution of energy or other
conserved quantity have been studied, the distribution of conserved
quantities is non Gaussian).

Hence, the relevant statistical ensemble for these models is the microcanonical
one. In the following, we will work only with microcanonical measures
as a base for all derivations. If equilibrium statistical mechanics
is relevant for slightly non-equilibrium situations in fluid mechanics
problems\textbf{,} it will most likely not be through canonical distributions,
but through microcanonical distributions.

In statistical mechanics studies, it is sometimes argued that, in
the limit of an infinite number of degrees of freedom, canonical and
microcanonical measures are equivalent. Thus, as canonical measures
are more easily handled, they are preferred in many works. However,
while the equivalence of canonical and microcanonical ensembles is
very natural and usually true in systems with short-range interactions
(commonly found in condensed matter theory), it is often not actually
so in systems with long-range interactions, such as the 2D Euler equations
(see for instance \cite{Bouchet_Barre:2005_JSP,Dauxois_Ruffo_Arimondo_Wilkens_2002LNP...602....1D,Campa_Dauxois_Ruffo_Revues_2009_PhR...480...57C,Bouchet_Gupta_Mukamel_PRL_2009,Chavanis_2006IJMPB_Revue_Auto_Gravitant,Bouchet:2008_Physica_D,EllisHavenTurkington:2000_Inequivalence}
and references therein).\\

In statistical mechanics, a macrostate $M$ is a set of microstates
verifying some conditions. The conditions are usually chosen such
that they describe conveniently the macroscopic behavior of the physical
systems through a reduced number of variables. For instance, in a
magnetic system, a macrostate $M$ could be the ensemble of microstates
with a given value of the total magnetization; in the case of a gas,
a macrostate could be the ensemble of microstates corresponding\textbf{
}to a given local density $f\left(\mathbf{x},\mathbf{p}\right)$ in
the six dimensional space $(\mathbf{x},\mathbf{p}$) ($\mu$ space),
where $f$ is defined  for instance through some coarse-graining.
In our fluid problem, an interesting macrostate will be the local
probability distribution $\rho\left(\mathbf{x},\sigma\right)d\sigma$
to observe vorticity values $\omega\left(\mathbf{x}\right)=\sigma$
at $\mathbf{x}$ with precision $\mathrm{d}\sigma$.

If we identify the macrostate $M$ with the values of the constraints
that define it, we can define the probability of a macrostate $P\left(M\right)dM$.
If the microstates are distributed according to the microcanonical
measure, $P\left(M\right)$ is proportional to the volume of the subset
$\Omega_{M}$ of phase space where microstates $\left\{ q_{i},p_{i}\right\} _{1\leq i\leq N}$
realize the state $M$. The Boltzmann entropy of a macrostate $M$
is then defined to be proportional to the logarithm of the phase space
volume of the subset $\Omega_{M}$ of all microstates $\left\{ q_{i},p_{i}\right\} _{1\leq i\leq N}$
that realize the state $M$.

In systems with a large number of degrees of freedom, it is customary
to observe that the probability of some macrostates is concentrated
close to a unique macrostate. There exist also cases where the probability
of macrostates concentrates close to larger set of macrostates (see
for instance \cite{Kiessling_2008AIPC}). Such a concentration is
a very important information about the macroscopic behavior of the
system. The aim of statistical physics is then to identify the physically
relevant macrostates, and to determine their probability and where
this probability is concentrated. This is the program we will follow
in the next sections, for the 2D Euler equations.

In the preceding discussion, we have explained that the microcanonical
measure is a natural invariant measure with given values of the invariants.
An important issue is to know if this measure describes also the statistics
of the temporal averages of the Hamiltonian system. This issue, called
ergodicity will be discussed in section 4.4.

\subsubsection{Hamiltonian structure for the 2D Euler equations\label{sub:Hamiltonian-structure-2D-Euler}}

The first step to define the microcanonical measure is to identify
the equivalent of a Liouville theorem and of the dynamical invariants.
The Euler equations describe a conservative dynamics. They can be
derived from a least-action principle \cite{Salmon_1998_Book,Holm_Marsden_Ratiu_1998_EulerPoincare},
just like canonical Hamiltonian systems. It is thus natural to expect
a Hamiltonian structure for them as well. There are however fundamental
differences between infinite-dimensional systems like the 2D Euler
equations and canonical Hamiltonian systems: 
\begin{enumerate}
\item The Euler equations consist in a dynamical system of infinite dimension.
The notion of volume of an infinite-dimensional space is meaningless.
Hence, the microcanonical measure cannot be defined straightforwardly. 
\item For such infinite-dimensional systems, we cannot in general find a
canonical structure (pair of canonically conjugated variables $\left\{ q_{i},p_{i}\right\} $
describing all degrees of freedom). It exists however a Poisson structure:
one can define a Poisson bracket $\left\{ \cdot\,,\cdot\right\} $,
like in canonical Hamiltonian systems (\ref{eq:Crochets_Poisson_Canonique}),
so that the dynamics writes \begin{equation}
\partial_{t}\omega=\left\{ \omega,\mathcal{H}[\omega]\right\} ,\label{eq:Poisson Bracket}\end{equation}
 where $\mathcal{H}$ is the Hamiltonian. 
\end{enumerate}
For infinite-dimensional Hamiltonian systems (such as the 2D Euler
equations), the Poisson bracket in (\ref{eq:Poisson Bracket}) is
often degenerate \cite{Holm_Marsden_Ratiu_Weinstein_1985PhysRev,Morrison_1998_HamiltonianFluid_RvMP},
leading to the existence of an infinite number of conserved quantities.
These conservation laws have very important dynamical consequences,
as explained in the next section. A detailed description of the Hamiltonian
structure of infinite-dimensional systems is beyond the scope of this
paper. We refer to \cite{Holm_Marsden_Ratiu_Weinstein_1985PhysRev,Morrison_1998_HamiltonianFluid_RvMP}
for a description of the Poisson and Hamiltonian structure for many
fluid systems. In the next two sections, the dynamical invariants
and the Liouville theorem are discussed in the context of the 2D Euler
equations.

\subsubsection{Casimir conservation laws\label{sub:Casimirs-conservation-laws}}

2D Euler equations \eqref{eq:Euler_2D_Vorticity} conserve an infinite
number of functionals, named Casimir invariants (or Casimirs for short).
They are all functionals of the form \begin{equation}
\mathcal{C}_{s}[\omega]=\int_{\mathcal{D}}\mathrm{d}\mathbf{r}\, s(\omega),\label{eq:casimir}\end{equation}
 where $s$ is any sufficiently smooth function. As mentioned in section
\ref{sub:Hamiltonian-structure}, Casimir conserved quantities are
related to the degeneracy of the Poisson structure in infinite-dimensional
Hamiltonian systems. They can also be understood as the invariants
arising from Noether's theorem, as a consequence of the relabeling
symmetry of fluid mechanics (see for instance \cite{Salmon_1998_Book}).

\smallskip{}
Let us note $A(\sigma)$ the area of $\mathcal{D}$ with vorticity
values less than $\sigma$, and $\gamma\left(\sigma\right)$ the vorticity
distribution: \begin{equation}
\gamma\left(\sigma\right)=\frac{1}{\left|\mathcal{D}\right|}\frac{\mathrm{d}A}{\mathrm{d}\sigma}\quad\mbox{with\,\,\,}A\left(\sigma\right)=\int_{\mathcal{D}}\mathrm{d}\mathbf{r}\,\chi_{\left\{ \omega\left({\bf x}\right)\leq\sigma\right\} }\,,\label{eq:distribution_vorticite}\end{equation}
where $\chi_{\mathcal{B}}$ is the characteristic function of the
set $\mathcal{B}\subset\mathcal{D}$, and $\left|\mathcal{D}\right|$
is the area of $\mathcal{D}$. Since equations \eqref{eq:Euler_2D_Vorticity}
express transport by an incompressible flow, the area $\gamma\left(\sigma\right)$
occupied by a given vorticity level $\sigma$ (or equivalently $A\left(\sigma\right)$)
is a dynamical invariant.

The conservation of the distribution $\gamma\left(\sigma\right)$
is equivalent to the conservation of all Casimir functionals (\ref{eq:casimir}).
The domain-averaged vorticity $\mathcal{G}$, enstrophy $\mathcal{G}_{2}$,
and higher moments of the vorticity $\{\mathcal{G}_{k}\}_{k\geq3}$
are Casimirs of particular interest: \begin{equation}
\text{for}\; k\geq1,\quad\mathcal{G}_{k}\left[\omega\right]=\int_{\mathcal{D}}\mathrm{d}\mathbf{r}\,\omega^{k}\qquad(\text{with}\;\;\mathcal{G}\left[\omega\right]:=\mathcal{G}_{1}\left[\omega\right]).\label{eq:Enstrophy}\end{equation}
 Note that if $\mathcal{D}$ is bounded, $\mathcal{G}$ is also the
circulation: $\mathcal{G}=\int_{\partial\mathcal{D}}\mathbf{v}\cdot\mathrm{d}\mathbf{l}\,.$

\medskip{}
In any Hamiltonian system, symmetries are associated with conservation
laws, as a consequence of Noether's theorem. Then, if the domain $\mathcal{D}$
is invariant under rotations or translations, there will be conservation
of angular momentum or linear momentum respectively. If the domain
displays such symmetries, these conservation laws have to be taken
into account in a statistical mechanics analysis.

\subsubsection{Detailed Liouville theorem and microcanonical measure for the dynamics
of conservative flows\label{sub:microcanonical-measure}}

In order to discuss the detailed Liouville theorem, and build the
microcanonical measure, we decompose the vorticity field on the eigenmodes
of the Laplacian on $\mathcal{D}$. We could decompose the field in
any other orthonormal basis. The Laplacian and Fourier bases prove
simpler for the following discussion, whereas finite-element bases
are much more natural to justify a mean-field approximation and to
obtain large-deviation results for the measures, as will be discussed
in section \ref{sub:Mean_Field}.

\medskip{}
We call $\{e_{i}\}_{i\geq1}$ the orthonormal family of eigenfunctions
of the Laplacian on $\mathcal{D}$: \begin{equation}
-\Delta e_{i}=\lambda_{i}e_{i}\,,\quad\int_{\mathcal{D}}\mathrm{d}\mathbf{r}\ e_{i}e_{j}=\delta_{ij}\,.\label{eq:LaplacianEigenmodes}\end{equation}
 The eigenvalues $\{\lambda_{i}\}$ are arranged in increasing order.
For instance, for a doubly periodic or infinite domain, $\{e_{i}\}$
are just the Fourier modes. Any function $g$ defined on $\mathcal{D}$
can be decomposed into $g(t,{\bf r})=\sum_{i}g_{i}(t)e_{i}(\mathbf{r})$
with $g_{i}(t)=\int_{\mathcal{D}}\mathrm{d}\mathbf{r}\ g(t,{\bf r})e_{i}({\bf r})$.
Then, \[
\omega({\bf r},t)=\sum_{i=1}^{+\infty}\omega_{i}(t)e_{i}({\bf r}).\]
From \eqref{eq:Euler_2D_Vorticity}, \begin{equation}
\dot{\omega}_{i}=A_{ijk}\omega_{j}\omega_{k}\,,\label{eq:Euler-Fourrier}\end{equation}
where the explicit expression of $A_{ijk}$ need not be known for
the following discussion. For (\ref{eq:Euler-Fourrier}), a detailed
Liouville theorem holds: \begin{equation}
\forall\, i,\quad\frac{\partial\dot{\omega_{i}}}{\partial\omega_{i}}=0\label{eq:Detailed_Liouville_Euler}
\end{equation}
(see \cite{Lee52}, \cite{Kraichnan_Motgommery_1980_Reports_Progress_Physics}).
Note that even though we have discussed the detailed Liouville theorem
in the context of mode decomposition, more general results exist \cite{Robert_2000_CommMathPhys-TruncationEuler}\cite{Zeitlin_1991_HamiltonianTruncations}
\footnote{A direct consequence of the detailed Liouville theorem (\ref{eq:Detailed_Liouville_Euler})
is that any truncation of the Euler equations also verifies a Liouville
theorem \cite{Kraichnan_Motgommery_1980_Reports_Progress_Physics}.
This result is actually much more general: any approximation of the
Euler equations obtained by an $L_{2}$-projection on a finite-dimensional
basis verifies a Liouville theorem (see \cite{Robert_2000_CommMathPhys-TruncationEuler}).
For truncations preserving the Hamiltonian structure and a finite
number of Casimir invariants, see \cite{Zeitlin_1991_HamiltonianTruncations}.%
}.

\smallskip{}

\paragraph*{Microcanonical measure}

From the detailed Liouville theorem (\ref{eq:Detailed_Liouville_Euler}),
we can define the microcanonical measure. First, let us define the
$n$-moment microcanonical measure: \begin{equation}
\mu_{m,n}\left(E,\Gamma_{1},\ldots,\Gamma_{n}\right)=\frac{1}{\Omega_{n}\left(E,\Gamma_{1},\ldots,\Gamma_{n}\right)}\prod_{i}\mathrm{d}\omega_{i}\ \delta\left(\mathcal{E}\left[\omega\right]-E\right)\prod_{k=1}^{n}\delta\left(\mathcal{G}_{k}\left[\omega\right]-\Gamma_{k}\right),\label{eq:microcanonical_measure_n}\end{equation}
 where $\mathcal{E}$ is the energy \eqref{eq:Energy} and $\{\Gamma_{k}\}$
are the vorticity moments \eqref{eq:Enstrophy}, the subscript $m$
still standing for `microcanonical'. A precise definition of
$\mu_{m,n}$ requires the definition of approximate finite-dimensional
measures: for any observable $\phi_{M}$ depending on $M$ components
$\{\omega_{i}\}_{1\leq i\leq M}$ of $\omega$, we define \[
\langle\mu_{m,n}^{N},\phi_{M}\rangle=\frac{1}{\Omega_{n,N}\left(E,\Gamma_{1},\ldots,\Gamma_{n}\right)}\int\prod_{i=1}^{N}\mathrm{d}\omega_{i}\ \delta\left(\mathcal{E}_{N}\left[\omega\right]-E\right)\prod_{k=1}^{n}\delta\left(\mathcal{G}_{k,N}\left[\omega\right]-\Gamma_{k}\right)\phi_{M}\left(\omega_{1},\ldots,\omega_{M}\right),\]
 where $\mathcal{E}_{N}$ and $\{\mathcal{G}_{k,N}\}$ are finite-dimensional
approximations of $\mathcal{E}$ \eqref{eq:Energy} and $\{\mathcal{G}_{k}\}$
\eqref{eq:Enstrophy} respectively. Then we define $\langle\mu_{m,n},\phi_{M}\rangle=\lim_{N\rightarrow\infty}\langle\mu_{m,n}^{N},\phi_{M}\rangle$.
As explained in the next paragraph, $\Omega_{n,N}$ has usually no
finite limit when $N$ goes to infinity, so the definition of the
normalization factor $\Omega_{n}\left(E,\Gamma_{1},\ldots,\Gamma_{n}\right)$
in the formal notation (\ref{eq:microcanonical_measure_n}) implies
a proper rescaling.

$\{\mu_{m,n}\}_{n\geq1}$ is expected to be a set of invariant measures
for the 2D Euler equations (this is easily verified through formal
computations). The microcanonical measure corresponding to the infinite
set of invariants $\left\{ \Gamma_{k}\right\} _{k\geq1}$ is then
defined by \[
\mu_{m}\left(E,\left\{ \Gamma_{k}\right\} \right)=\lim_{n\rightarrow\infty}\mu_{m,n}\left(E,\Gamma_{1},\ldots,\Gamma_{n}\right),\]
 and is denoted by \begin{equation}
\mu_{m}\left(E,\left\{ \Gamma_{k}\right\} \right)=\frac{1}{\Omega\left(E,\left\{ \Gamma_{k}\right\} \right)}\prod_{i=1}^{\infty}\mathrm{d}\omega_{i}\ \delta\left(\mathcal{E}\left[\omega\right]-E\right)\prod_{k=1}^{\infty}\delta\left(\mathcal{G}_{k}\left[\omega\right]-\Gamma_{k}\right).\label{eq:microcanonical_measure}\end{equation}

\smallskip{}

\paragraph*{Equilibrium Boltzmann entropy}

The normalization factors $\Omega_{n}\left(E,\Gamma_{1},\ldots,\Gamma_{n}\right)$
and $\Omega\left(E,\left\{ \Gamma_{k}\right\} \right)$ define the
Boltzmann entropies \begin{equation}
S_{n}\left(E,\Gamma_{1},\ldots,\Gamma_{n}\right)=k_{B}\log\Omega_{n}\left(E,\Gamma_{1},\ldots,\Gamma_{n}\right)\,\,\,\mbox{and}\,\,\, S\left(E,\left\{ \Gamma_{k}\right\} \right)=k_{B}\log\Omega\left(E,\left\{ \Gamma_{k}\right\} \right).\label{eq:Entropy_Definition_Ini}\end{equation}
 The behavior of expressions like $\int\prod_{i=1}^{N}\mathrm{d}\omega_{i}\ \delta\left(\mathcal{E}_{N}\left[\omega\right]-E\right)\prod_{k=1}^{n}\delta\left(\mathcal{G}_{k,N}\left[\omega\right]-\Gamma_{k}\right)$,
for large $N$, is expected to be typically of the form $C(N)\exp\left(NS_{n}\left(E,\Gamma_{1},\ldots,\Gamma_{n}\right)\right)$.
In the definition of $\Omega_{n}$, the prefactor $C(N)$ is omitted
(the entropy is defined up to a constant independent of the physical
variables), so $S_{n}$ and $S$ are actually `specific entropies'
(entropies per degree of freedom).

\subsection{Validity of a mean-field approach to the microcanonical measures\label{sub:Valid_Mean_Field}}

In the previous section, we defined the microcanonical measure for
the 2D Euler equations. In this section, we give a heuristic explanation
of the reason why a mean-field description of the microcanonical measure
is exact and give references for more precise results (large deviations
for sets of measures). At the core of our discussion about mean-field
approaches, lies the result that\textbf{ }if the microstates are distributed
according to the microcanonical measure, the probability distributions
of vorticity at different points are independent (product measure).
This is not only a result of precise mathematical works (large deviations
for sets of measures), but also the deep reason for the validity of
the mean-field approach. This is also the main reason for the interest
of Young measures for the 2D Euler equations \cite{Michel_Robert_LargeDeviations1994CMaPh.159..195M}.

Because the large deviations of sets of measures are rather technical
results from probability theory, for pedagogic reasons, we study the
energy-enstrophy microcanonical ensemble in section \ref{sub:Direct_Comp_energie-enstrophie},
with elementary mathematical tools. We prove in Appendix D that the
correlation coefficient between $\omega(\mathbf{r})$ and $\omega(\mathbf{r'})$
is zero for the energy-enstrophy microcanonical measure. It could
be proved without much difficulty that, in addition, $\omega(\mathbf{r})$
and $\omega(\mathbf{r'})$ are actually independent variables. As
said above, the statistical independence of vorticity values at different
points is a much more general result and is essential. Let us first
analyze an extremely important implication: the possibility to quantify
the phase space volume (Boltzmann entropy) through the Boltzmann--Gibbs
formula.

\smallskip{}

\paragraph*{Boltzmann entropy of a macrostate and Boltzmann--Gibbs formula}

A classical example where degrees of freedom can be considered independent
is an ensemble of particles (say, hard spheres) undergoing collisions
in the dilute limit (Boltzmann--Grad limit~\cite{Spohn_1991}). Microscopically,
particles travel with typical velocity $\bar{v}$ and collide with
each other after traveling a typical distance $l$, called the mean
free path. Let $\sigma$ be the diffusion cross-section for these
collisions. One has $\sigma=\pi a^{2}$, where parameter $a$ is of
the order of the particle radius. The mean free path is defined as
$l=1/\!(\pi a^{2}n)$, where $n$ is the typical particle density.
The Boltzmann equation applies when the ratio $a/l$ is small (Boltzmann--Grad
limit). In the limit $a/l\to0\,,$ any two colliding particles can
be considered independent (uncorrelated) as they come from very distant
regions. This is the base of Boltzmann's hypothesis of molecular chaos
(Stosszahl Ansatz). It explains why the evolution of the $\mu$-space
distribution function $f(\mathbf{x},\mathbf{p},t)$ may be described
by an autonomous equation, the Boltzmann equation (the $\mu$-space
is the six-dimensional space of spatial variable $\mathbf{x}$ and
momentum $\mathbf{p}$). In statistical mechanics, a macrostate
$M$ is a set of microstates verifying some conditions. The conditions
are usually chosen such that they describe conveniently the macroscopic
behavior of the physical system through a reduced number of variables.
The Boltzmann entropy of a macrostate $M$ is defined to be proportional
to the logarithm of the phase space volume of the subset $\Omega_{M}$
of all microstates $\left\{ q_{i},p_{i}\right\} _{1\leq i\leq N}$
that realize the state $M$. In the case of a dilute gas, the distribution
$f\left(\mathbf{x},\mathbf{p}\right)$ can be identified with the
macrostate: it is the set of all possible microstates $\left\{ \mathbf{x}_{i},\mathbf{p}_{i}\right\} _{1\leq i\leq N}$
such that the number of particles in the volume element $\Delta\mathbf{x}\Delta\mathbf{p}$
around $\left(\mathbf{x},\mathbf{p}\right)$ is $f\left(\mathbf{x},\mathbf{p}\right)\Delta\mathbf{x}\Delta\mathbf{p}$
(a precise mathematical definition goes through the limit $N\rightarrow\infty$,
see \cite{Goldstein_Lebowitz_2004PhyD..193...53G}). We note that
the Boltzmann entropy of the subset of phase space with fixed invariants
is the equilibrium Boltzmann entropy defined in section \ref{sub:microcanonical-measure},
formula (\ref{eq:Entropy_Definition_Ini}).

There is a classical argument by Boltzmann (which can be found in
any good textbook on statistical mechanics) to prove that the Boltzmann
entropy of the distribution $f$ is, up to a multiplicative constant,
given by the Boltzmann--Gibbs formula:
\begin{equation}
\mathcal{S}[f]=-\int\mathrm{d}\mathbf{x}\mathrm{d}\mathbf{p}\, f\log f.\label{eq:Maxwell-Boltzmann-Entropy}\end{equation}
We stress that this formula for the Boltzmann entropy is not a Gibbs
entropy %
\footnote{The Gibbs entropy $S=-k\int\rho(p_{i},q_{i})\log_{2}(\rho(p_{i},q_{i}))\, dp_{i}dq_{i}$
is an ensemble entropy, a weight on the phase space, whereas the Boltzmann--Gibbs
entropy is an integral over the $\mu$-space. In the case of dilute
gases, the Boltzmann--Gibbs entropy is just the opposite of the $H$
function of Boltzmann. We avoid this terminology here since our discussion
is not related to relaxation towards equilibrium, and because the
equivalent of an $H$ theorem has never been proved for the 2D Euler
equations.%
}. The essential point is that formula \eqref{eq:Maxwell-Boltzmann-Entropy}
is a valid counting of the volume of the accessible part of phase
space, only when particles can be considered independent. For instance,
for particles with short-range interactions studied by Boltzmann,
this is valid only in the Boltzmann--Grad limit.

As discussed above, for the microcanonical measure of the 2D Euler
equations, vorticity field values are independent. As we will explain
below, the reason is completely different from the Boltzmann case:
there is now no dilute-gas (Boltzmann--Grad) limit. Nevertheless,
the consequences will be the same: if we define $\rho\left(\mathbf{r},\sigma\right)$
such that $\rho\left(\mathbf{r},\sigma\right)\mathrm{d\mathbf{r}}\mathrm{d}\sigma$
be the probability to have values of $\omega$ between $\sigma$ and
$\sigma+\mathrm{d}\sigma$ in the area element $\mathrm{d}\mathbf{r}$
around $\mathbf{r}$, then the entropy \begin{equation}
\mathcal{S}[\rho]=-\int_{\mathcal{D}}\mathrm{d}\mathbf{r}\int_{-\infty}^{+\infty}\mathrm{d}\sigma\,\rho\ln\rho,\label{eq:Maxwell-Boltzmann-Entropy-Euler}
\end{equation}
actually quantifies the phase space volume. In order to give a precise
meaning of this last sentence, we first define the mean-field microcanonical
variational problem.

\smallskip{}

\paragraph*{Mean-field microcanonical variational problem}

As $\rho$ is a local probability, it verifies a local normalization
\begin{equation}
N\left[\rho\right](\mathbf{r})\equiv\int_{-\infty}^{+\infty}\mathrm{d}\sigma\,\,\rho\left(\sigma,\mathbf{r}\right)=1.\label{eq:normalisation}
\end{equation}
The average vorticity, for probability density $\rho$, is \begin{equation}
\bar{\omega}\left(\mathbf{r}\right)=\int_{-\infty}^{+\infty}\ \mathrm{d}\sigma\,\sigma\rho\left(\sigma,\mathbf{r}\right).\label{eq:vorticite_coarse_grained}
\end{equation}
The average vorticity \eqref{eq:vorticite_coarse_grained} is related
to the average streamfunction $\bar{\psi}$ so that $\bar{\omega}=\Delta\bar{\psi}$.

The conservation of all Casimir functionals \eqref{eq:casimir}, or
equivalently of the known vorticity distribution \eqref{eq:distribution_vorticite},
imposes a constraint on the local probability density $\rho$: \begin{equation}
D\left[\rho\right](\sigma)\equiv\int_{\mathcal{D}}\mathrm{d}\mathbf{r}\,\rho\left(\sigma,{\bf \mathbf{r}}\right)=\gamma\left(\sigma\right).\label{eq:distribution_ro}\end{equation}
 Then the mean-field entropy $S$ of the system is given by the variational
problem \begin{equation}
\tag{MVP}S(E,\gamma)=\sup_{\{\rho\,|\, N[\rho]=1\}}\left\{ \mathcal{S}[\rho]\ |\ \mathcal{E}\left[\bar{\omega}\right]=E,\, D\left[\rho\right]=\gamma\ \right\} .\label{eq:MVP}\end{equation}
 where $\mathcal{E}\left[\bar{\omega}\right]$ is the energy \eqref{eq:Energy}
of the average vorticity field $\bar{\omega}$.

An essential point has to be noted about the energy constraint in
\eqref{eq:MVP}: the constraint is expressed in terms of the average
vorticity field (hence the expression, `mean-field approximation'),
meaning that correlations between vorticity values at different points
are negligible, and meaning also that when computing the energy, fluctuations
around the average may be neglected.

\medskip{}
An essential point is that, up to addition of constant terms (i.e.,
independent of the physical parameters), the mean-field entropy \eqref{eq:MVP}
is exactly the same as the Boltzmann entropy defined from the rescaled
logarithm of the phase space volume, in equation \eqref{eq:Entropy_Definition_Ini}.
The definition of the entropy \eqref{eq:Entropy_Definition_Ini} and
the variational problem \eqref{eq:MVP} seem so different, that the
fact that they express the same concept is astonishing. This type
of results is indeed one of the great achievements of statistical
mechanics. In section \ref{sub:Direct_Comp_energie-enstrophie}, we
show that it is verified in the case of the energy-enstrophy measure,
using explicit elementary computations.

\smallskip{}

\paragraph*{Why is the mean-field entropy equal to the Boltzmann entropy?}

The deep reason why vorticity field values are independent for microcanonical
measures, and henceforth why entropy can be expressed by (\ref{eq:Entropie_Equilibre_MeanField_Energie_Enstrophie})
can be explained rather easily at a heuristic level. Correlations
between variables could appear through the dynamical constraints only:
energy, Casimirs, and so on. For instance, the energy of the 2D Euler
system can be expressed in a form where interactions between vorticity
values appear explicitly, using the Laplacian Green function $H\left(\mathbf{r},\mathbf{r}^{\prime}\right)$
(defined by $\Delta H\left(\mathbf{r},\mathbf{r}^{\prime}\right)=\delta(\mathbf{r},\mathbf{r}^{\prime}$)
with Dirichlet boundary conditions), we have: \begin{equation}
\mathcal{E}[\omega]=-\frac{1}{2}\int_{\mathcal{D}}\mathrm{d}\mathbf{r}\,\omega(\Delta^{-1}\omega)=-\frac{1}{2}\int_{\mathcal{D}}\int_{\mathcal{D}}\mathrm{d}\mathbf{r}\mathrm{d}\mathbf{r}'\,\omega\left(\mathbf{r}\right)H(\mathbf{r},\mathbf{r}^{\prime})\omega({\bf r'}).\label{eq:EnergyGreen}\end{equation}
 In formula (\ref{eq:EnergyGreen}) above, $H\left(\mathbf{r},\mathbf{r}^{\prime}\right)$
appears as the coupling between vorticity at point $\mathbf{r}$ and
vorticity at point $\mathbf{r}'$. The Laplacian Green function in
a two-dimensional space is logarithmic, hence non-local. Thus,
$\omega(\mathbf{r})$ is coupled to the vorticity at any other point
of the domain, not only close points.

For people trained in statistical mechanics, it is natural in systems
where degrees of freedom are coupled to many others, to consider these
degrees of freedom statistically independent at leading order, and
a mean-field approach should be a valid approximation. For example,
in systems with nearest-neighbor interactions, a mean-field approach
becomes exact in high dimensions, when the effective number of degrees
of freedom to which one degree of freedom is coupled becomes infinite.
For people not trained in statistical mechanics, this can be understood
simply: when the number of coupled degrees of freedom increases to
infinity, the interaction felt by one degree of freedom is no more
sensitive to the fluctuations of the others, but just to their average
value, owing to an effect similar to what happens for the law of large
numbers. Then a mean-field treatment becomes exact, which is equivalent
to saying that different degrees of freedom may be considered statistically
independent.

Because of the non-locality of the Green function, the vorticity field
at one point is virtually coupled to an infinite number of degrees
of freedom, and then a mean-field treatment is exact. This also explains
why the energy appearing in the variational problem (\ref{eq:MVP})
is computed from the average vorticity field.

\medskip{}
To formalize the preceding heuristic explanation, in order to prove
that the mean-field approximation is exact and that the Boltzmann--Gibbs
formula (\ref{eq:Maxwell-Boltzmann-Entropy}) is relevant, we need
a rather technical discussion. We will not explain this in detail.
This was justified by theoretical physicists for the point-vortex
model in the seventies (assumed to be valid by Joyce and Montgomery
\cite{Joyce_Montgommery_1973} and later proved to be self-consistent
in a Kramer-Moyal expansion). In the eighties, rigorous mathematical
proofs were given also for the point-vortex model (see \cite{Eyink_Spohn_1993_JSP....70..833E,Kiessling_Lebowitz_1997_PointVortex_Inequivalence_LMathPhys,CagliotiLMP:1995_CMP_II(Inequivalence)}
and references therein). In the modern formulation of statistical
mechanics, the entropy appears as a large-deviation rate function
for an ensemble of measures, justifying (\ref{eq:Maxwell-Boltzmann-Entropy})
and the variational problem (\ref{eq:MVP}). The proof of such large-deviation
results leading to the microcanonical measure for the 2D Euler equations,
justifying the mean-field approach, can be found in \cite{Michel_Robert_LargeDeviations1994CMaPh.159..195M}
(see also \cite{Boucher_Ellis_1999_AP} and references therein).\\

We thus conclude that a mean-field approach to the microcanonical
measure of the 2D Euler equations is valid. This justifies the use
of entropy (\ref{eq:Maxwell-Boltzmann-Entropy-Euler}) and of the
variational problem (\ref{eq:MVP}). This step is a crucial one as
it leads to a drastic simplification compared to a direct computation
from the definition of the microcanonical measure
\eqref{eq:microcanonical_measure_n}--\eqref{eq:microcanonical_measure}.
The first presentation of the equilibrium statistical mechanics of
the 2D Euler equations in this form dates from the beginning of the
`90s with the works of Robert and Sommeria, and those of Miller
\cite{Robert:1990_CRAS,Miller:1990_PRL_Meca_Stat,Robert:1991_JSP_Meca_Stat,RobertSommeria:1992_PRL_Relaxation_Meca_Stat}.
Thus, we call this theory the Robert--Sommeria---Miller (RSM) theory.

\subsection{Solutions to the mean-field variational problem for the microcanonical
measure\label{sub:Mean_Field}}

The aim of this section is to describe the critical points of the
mean-field variational problem (\ref{eq:MVP}), following the first
papers \cite{Robert:1990_CRAS,Miller:1990_PRL_Meca_Stat,Robert:1991_JSP_Meca_Stat,RobertSommeria:1992_PRL_Relaxation_Meca_Stat}.
For this purpose, we use the Lagrange multiplier rule to take account
of the constraints: the first variations of (\ref{eq:MVP}), \[
\delta\mathcal{S}-\int_{\mathcal{D}}\mathrm{d}{\bf r}\, A({\bf r})\delta N({\bf r})-\beta\delta\mathcal{E}-\int_{-\infty}^{+\infty}\,\mathrm{d\sigma}\alpha(\sigma)\delta\gamma(\sigma)=0\]
 are zero for any perturbation $\delta\rho$, where $A({\bf r}),\,\beta$,
and $\alpha(\sigma)$ are the Lagrange multipliers associated with
the conservation of $N({\bf r})$, $\mathcal{E}$ and $\gamma(\sigma)$
respectively. We obtain that the probability density distribution
$\rho$ verifies the Gibbs state equation: \begin{equation}
\rho\left(\sigma,\mathbf{r}\right)=\frac{e^{\beta\sigma\bar{\psi}\left(\mathbf{r}\right)-\alpha(\sigma)}}{Z_{\alpha}\left(\beta\bar{\psi}\left(\mathbf{r}\right)\right)}\quad\mbox{with}\,\,\, Z_{\alpha}\left(u\right)=\int_{-\infty}^{+\infty}\mathrm{d}\sigma\,\exp\left(\sigma u-\alpha\left(\sigma\right)\right).\label{eq:Gibbs-States}\end{equation}
 We see that $\rho$ depends on $\mathbf{r}$ through the average
streamfunction $\bar{\psi}$ only. From (\ref{eq:vorticite_coarse_grained})
and (\ref{eq:Gibbs-States}), we see that there is a functional relation
between the equilibrium average vorticity and the streamfunction:
\begin{equation}
\bar{\omega}=g\left(\beta\bar{\psi}\right)\,\,\,\mbox{with}\,\,\, g\left(u\right)=\frac{\mathrm{d}}{\mathrm{d}u}\log Z_{\alpha}(u).\label{eq:q-psi equilibre}\end{equation}
 This last equation characterizes the statistical equilibrium. It
should be solved for any value of $\left(\beta,\alpha(\sigma)\right)$.
Then, one has to compute the energy and vorticity distributions as
functions of $\beta$ and $\alpha(\sigma)$. For given energy $E$
and distribution $\gamma(\sigma)$, among all possible values of $\left(\beta,\alpha(\sigma),\rho(\sigma)\right)$
solving (\ref{eq:Gibbs-States}-\ref{eq:q-psi equilibre}), the maximizer
of the entropy (\ref{eq:MVP}) is selected.

\section{Energy-enstrophy microcanonical measure for the 2D Euler equations\label{sec:Energy-Enstrophy-microcan-meas}}

\noindent The energy-enstrophy microcanonical measure is defined as
\begin{equation}
\mu_{m,K}\left(E,\Gamma_{2}\right)=\frac{1}{\Omega\left(E,\Gamma_{2}\right)}\prod_{i=1}^{\infty}\mathrm{d}\omega_{i}\,\delta(\mathcal{E}\left[\omega\right]-E)\delta(\mathcal{G}_{2}\left[\omega\right]-{\Gamma}_{2}).\label{eq:Energy-Enstophy-Mu}\end{equation}
 This is the measure where only the quadratic invariants are taken
into account. There is \emph{a priori} no physical reason to exclude
the other invariants; however, the energy-enstrophy microcanonical
measure can be interesting, because it is, in some cases, a good approximation
of the complete microcanonical measure. Our real motivation to treat
it in detail is rather pedagogical: it will be very useful to prove
with this simple example, using elementary explicit computation, the
equivalence between the microcanonical measure introduced in section
\ref{sub:microcanonical-measure} through Fourier mode decomposition,
and the solution to the microcanonical mean-field variational problem
of section \ref{sub:Mean_Field}.

\noindent We compute the entropy and the probability distribution
function for the amplitude of each mode. These computations are performed,
always in the microcanonical ensemble, on one hand directly from the
definition of the energy-enstrophy measure (section \ref{sub:Direct_Comp_energie-enstrophie}),
and on the other hand from the mean-field variational problem (section
\ref{sub:En-En-MF}).

The energy-enstrophy measure was treated and discussed at length by
many authors in the `70s, including Kraichnan (see \cite{Kraichnan_Motgommery_1980_Reports_Progress_Physics},
a precise discussion can be found in\textbf{ }\cite{Majda_Wang_Book_Geophysique_Stat}).
However, these computations were always performed in the canonical
ensemble. The following discussion gives the first derivation in the
microcanonical ensemble, and the first observation of ensemble inequivalence
for the energy-enstrophy ensembles --- microcanonical and canonical
(section \ref{sub:Ensemble-inequivalence-Energy-Enstrophie}). The
energy-enstrophy ensembles are an elementary example of the so-called
partial equivalence\textbf{ }\cite{EllisHavenTurkington:2000_Inequivalence},
in the theory of ensemble inequivalence. We\textbf{ }shall come back
to discuss Kraichnan-type results in section \ref{sub:Kraichnan}.

\subsection{Direct computation of the energy-enstrophy measure from its finite-dimensional
approximation\label{sub:Direct_Comp_energie-enstrophie}}

Following the discussion of section \ref{sub:microcanonical-measure},
the energy-enstrophy microcanonical measure is defined through $N$-dimensional
approximations: \begin{equation}
\mu_{m,K}=\lim_{N\rightarrow\infty}\mu_{m,K}^{N}\,\quad\mbox{with}\quad\mu_{m,K}^{N}=\frac{1}{\Omega_{K,N}\left(E,\Gamma_{2}\right)}\prod_{i=1}^{N}\mathrm{d}\omega_{i}\,\delta(\mathcal{E}_{N}\left[\omega\right]-E)\delta(\mathcal{G}_{2,N}\left[\omega\right]-\Gamma_{2}),\label{eq:microcanonical_energy_enstrophy_N}\end{equation}
 where we use the same notation as in section \ref{sub:microcanonical-measure},
so $2\mathcal{E}_{N}\left[\omega\right]=\sum_{i=1}^{N}\omega_{i}^{2}/\lambda_{i}$
and ${\Gamma}_{2,N}\left[\omega\right]=\sum_{i=1}^{N}\omega_{i}^{2}$.
In the following, we assume that the first mode is non-degenerate:
$\lambda_{1}\neq\lambda_{2}$ (this is always true for simply connected
bounded Lipschitz domains, but this is wrong for doubly periodic boundary
conditions in a square domain).

\smallskip{}
The main technical difficulty is to compute \begin{equation}
\Omega_{K,N}\left(E,\Gamma_{2}\right)=\int\prod_{i=1}^{N}\mathrm{d}\omega_{i}\,\delta(\mathcal{E}_{N}\left[\omega\right]-E)\delta(\mathcal{G}_{2,N}\left[\omega\right]-{\Gamma}_{2}),\label{eq:Omega_N}\end{equation}
and the entropy \begin{equation}
S_{K}\left(E,\Gamma_{2}\right)=\lim_{N\rightarrow\infty}\frac{1}{N}\log\left[\Omega_{K,N}\left(E,\Gamma_{2}\right)\right]-C(N,\left\{ \lambda_{i}\right\}),\label{eq:Definition-Entropy}
\end{equation}
 where $C$ does not depend on the physical parameters. It depends
only on $N$ and on the geometric factors $\left\{ \lambda_{i}\right\} $,
and can be discarded as the entropy is always defined up to an arbitrary
constant.

The computation of $\Omega_{K,N}$ and $S_{K}$, using representation
of the delta function as an integral in the complex plane, is given
in Appendix B. It yields the result \begin{align}
\Omega_{K,N}\left(E,\Gamma_{2}\right) & \underset{N\rightarrow\infty}{\sim}C_{3}\left(N,\left\{ \lambda_{i}\right\} \right)C_{4}\left(\left\{ \lambda_{i}\right\} ,\Gamma_{2},N\right)\frac{\exp\left[NS_{K}\left(E,\Gamma_{2}\right)\right]}{\sqrt{2E}}\left[1+o\left(\frac{1}{N}\right)\right]\,,\notag\label{eq:Entropie-energie-enstrophie-texte}\\
\mbox{with}\quad S_{K}\left(E,\Gamma_{2}\right) & =\frac{1}{2}\log\left(\Gamma_{2}-2\lambda_{1}E\right)+\frac{\log2}{2}\,,\end{align}
 (see \eqref{eq:Volume_espace_phases_energie_enstrophie} and \eqref{eq:Entropie-Energie-Enstrophie},
page \pageref{eq:Volume_espace_phases_energie_enstrophie}), where
$C_{3}\left(N,\left\{ \lambda_{i}\right\} \right)$ does not depend
on the energy or enstrophy, and $C_{4}$ has no exponentially large
contribution ($\lim_{N\rightarrow\infty}\left(\log C_{4}\right)/N=0$).
\\

We now describe finite-$N$ effects for $\mu_{m,K}^{N}$ \eqref{eq:microcanonical_energy_enstrophy_N},
the finite-$N$ approximation of the energy-enstrophy microcanonical
measure\textbf{ $\mu_{m,K}$} (\ref{eq:Energy-Enstophy-Mu}). It is
easy to see from \eqref{eq:microcanonical_energy_enstrophy_N}, that
for the $N$-dimensional measure $\mu_{m,K}^{N}$, the distribution
function for the amplitude $\omega_{n}$ of mode $e_{n}$ is given
by \begin{equation}
P_{N,n}\left(\omega_{n}\right)=\frac{\Omega_{K,N-1;\lambda_{n}}\left(E-\omega_{n}^{2}/2\lambda_{n},\Gamma_{2}-\omega_{n}^{2}\right)}{\Omega_{K,N}\left(E,\Gamma_{2}\right)},\label{eq:PDF-Omega-n}\end{equation}
 where the definition of $\Omega_{K,N-1;\lambda_{n}}$ is the same
as that of\textbf{ $\Omega_{K,N}$ \eqref{eq:Omega_N}}, but with
integration over $\omega_{n}$ excluded, and with constraint $\omega_{n}^{2}\leq\max\left\{ 2\lambda_{n}E,\Gamma_{2}\right\} $.

The distribution function for the energy $E_{n}=\omega_{n}^{2}/2\lambda_{n}$
of mode $e_{n}$ is obtained through the change of variable $P_{N,n}\left(E_{n}\right)\mathrm{d}E_{n}=P_{N,n}\left(\omega_{n}\right)\mathrm{d}\omega_{n}$.
Using result (\ref{eq:Entropie-energie-enstrophie-texte}) for both
$\Omega_{K,N-1;\lambda_{1}}$ (then $\lambda_{1}$ has to be replaced
with $\lambda_{2}$) and $\Omega_{K,N}$, we obtain \begin{equation}
P_{N,1}\left(E_{1}\right)\underset{N\rightarrow\infty}{\sim}C\,\frac{\exp\left[N\log\left(\Gamma_{2}-2\lambda_{2}E+2\left(\lambda_{2}-\lambda_{1}\right)E_{1}\right)/2\right]}{\sqrt{E_{1}\left(E-E_{1}\right)}}\quad\mbox{for}\;\;0<E_{1}<E,\label{eq:Larges-Deviations-E1}\end{equation}
 and $P_{N,1}\left(E_{1}\right)=0$ otherwise. $C$ is a normalization
constant which does not depend on $E_{1}$. From this expression,
we can see that the most probable energy is $E_{1}=E$. Moreover,
the distribution is exponentially peaked around $E_{1}=E$, so that
in the infinite-$N$ limit (energy-enstrophy microcanonical distribution),
we have \[
P_{1}\left(E_{1}\right)=\delta(E-E_{1}).\]
 This is a striking result: for the energy-enstrophy microcanonical
measure, all the energy condensates in the first mode.\\

The type of result (\ref{eq:Larges-Deviations-E1}) is called a large-deviation
result. It describes accurately the distribution function for $E_{1}$,
however large the deviations from the most probable value, in the
limit $N\rightarrow\infty$. If large deviations in $E-E_{1}$ are
disregarded, a good approximation for large $N$ of the finite-$N$
distribution is the exponential distribution: \begin{equation}
P_{N,1}\left(E_{1}\right)=\underset{N\rightarrow\infty}{\sim}C\,\frac{\exp\left[-N\frac{\lambda_{2}-\lambda_{1}}{\Gamma_{2}-2\lambda_{1}E}\left(E-E_{1}\right)\right]}{\sqrt{E-E_{1}}}\quad\mbox{for}\;\;0<E_{1}<E\;\;\mbox{and}\;\; N^{1/2}\left(E_{1}-E\right)\ll1.\label{eq:PE1}\end{equation}
 We note that the distribution for $\omega_{1}$ is exponential as
well. The amplitude of the departure of $E_{1}$ from value $E$ is
thus proportional to $1/N$ and to $\left(\Gamma_{2}-2\lambda_{1}E\right)/\left(\lambda_{2}-\lambda_{1}\right).$

The distribution of the energy $E_{n}$ of mode $e_{n}$$ $ is obtained
similarly as \begin{equation}
P_{N,n}\left(E_{n}\right)\underset{N\rightarrow\infty}{\sim}C\frac{\exp\left[N\log\left(\Gamma_{2}-2\lambda_{1}E-2\left(\lambda_{n}-\lambda_{1}\right)E_{n}\right)\right]}{\sqrt{E_{n}}}\,\,\,\mbox{for}\,\,\,0\leq E_{n}\leq E.\label{eq:PEn}\end{equation}
 For infinite $N$, the energy-enstrophy microcanonical distribution
is thus a delta function with zero energy: \[
P_{n}\left(E_{n}\right)=\delta\left(E_{n}\right).\]
 Disregarding large deviations, the finite-$N$ distribution is also
well approximated by an exponential distribution (this time a Gaussian
distribution for $\omega_{n}$), with typical departure from $0$
of order $1/N$ for the energy and a variance of order $1/\sqrt{N}$
for $\omega_{n}$. It may also be checked that for large $n$ ($\lambda_{n}\gg\lambda_{1}$),
the variance of the enstrophy becomes independent of $n$ (asymptotic
equipartition of the enstrophy).

Results such as \eqref{eq:Larges-Deviations-E1}--\eqref{eq:PEn} are
classical in statistical mechanics: typical departures from the most
probable value have a Gaussian distribution with variance of order
$1/\sqrt{N}$, except for variables whose most probable values are
at the edge of the accessible range. In the latter case (for instance,
$E_{n}$ or $\omega_{1}$ in the example discussed above), the distribution
is exponential with typical departure of order $1/N$.

We note that for the 2D Euler equations, only the infinite-$N$ limit
is relevant, and finite-$N$ effects have no dynamical counterpart.
They may be of interest for truncated systems only.\\

From the preceding discussion, we see that all the energy is concentrated
in the first mode, and that the excess enstrophy $\Gamma_{2}-2\lambda_{1}E$
goes to smaller an smaller scales, leading to zero energy and zero
enstrophy in every mode except the first one. This condensation of
energy in the first mode is the main physical prediction of the microcanonical
energy-enstrophy ensemble.

\subsection{Energy-enstrophy microcanonical measure from a mean-field approach\label{sub:En-En-MF}}

Let us compute the entropy in the (microcanonical) energy-enstrophy
ensemble, now starting from the mean-field variational problem, and
compare the results with those of section \ref{sub:Direct_Comp_energie-enstrophie}.

The mean-field variational problem in the (microcanonical) energy-enstrophy
ensemble is the equivalent of (\ref{eq:MVP}) but with only quadratic
invariants taken into account: \begin{equation}
S_{K}\left(E,\Gamma_{2}\right)=\sup_{\left\{ \rho|N\left[\rho\right]=1\right\}} \left\{ \frac{1}{\mathcal{\left|D\right|}}\mathcal{S}[\rho]\ |\ \mathcal{E}\left[\overline{\omega}\right]=E\ ,\,\int\mathrm{d}\mathbf{r}\mathrm{d}\sigma\,\sigma^{2}\rho=\Gamma_{2}\right\} \label{eq:Entropie_Equilibre_MeanField_Energie_Enstrophie}\end{equation}

Note that we seek here to maximize the specific entropy $\mathcal{S}[\rho]/|\mathcal{D}|$,
because this is what leads to the actual measure of phase space volume.
It is customary in the literature to ignore the $1/|\mathcal{D}|$
prefactor for convenience, as done for instance in \eqref{eq:MVP}.

In order to compute the critical points of the constrained variational
problem \eqref{eq:Entropie_Equilibre_MeanField_Energie_Enstrophie},
we introduce Lagrange multipliers $A({\bf r}),\beta$, and $\alpha$,
associated with the conservation of $N({\bf r}),\mathcal{E}$, %$N\left(\mathbf{r}\right)$,
and $\mathcal{G}_{2}$ respectively (see \eqref{eq:normalisation},
\eqref{eq:Energy} and \eqref{eq:Enstrophy}, respectively, for the
expression of these quantities). Critical points of \eqref{eq:Entropie_Equilibre_MeanField_Energie_Enstrophie}
are such that \[
\frac{\delta\mathcal{S}}{|{\mathcal{D}}|}-\int_{\mathcal{D}}\mathrm{d}{\bf r}\, A({\bf r})\delta N({\bf r})-\beta\delta\mathcal{E}-\alpha\delta\mathcal{G}_{2}=0\quad\forall\;\delta\rho.\]
 This is equivalent to \[
\rho(\sigma,\mathbf{r)}=\rho^{*}(\sigma,{\bf r})\equiv B({\bf r})\mathrm{e}^{|{\mathcal{D}}|(\beta\sigma\bar{\psi}({\bf r})-\alpha\sigma^{2})},\]
 %-\frac{1+\log\rho}{\mathcal{D}}-A({\bf r})+\beta\sigma\bar{\psi}-\alpha\sigma^{2}=0\quad\Rightarrow\quad
where the prefactor $B({\bf r})\equiv\exp(-1-A({\bf r})|{\mathcal{D}}|)$
is determined from the normalization constraint: \begin{equation}
\rho^{*}(\sigma,{\bf r})=\sqrt{\frac{\alpha|{\mathcal{D}}|}{\pi}}\mathrm{e}^{-\alpha|{\mathcal{D}}|\left(\sigma-\frac{\beta\bar{\psi}({\bf r})}{2\alpha}\right)^{2}}.\label{eq:en-en-rho-crit}\end{equation}
 The computation above, yielding a Gaussian distribution in the energy-enstrophy
ensemble, is a classical result noted in many previous works (see,
for instance, \cite{Chavanis_Sommeria_1998JFM_LocalizedEquilibria...356..259C}).

Substituting expression (\ref{eq:en-en-rho-crit}) into \eqref{eq:Maxwell-Boltzmann-Entropy-Euler},
we get \begin{equation}
\frac{1}{\mathcal{\left|D\right|}}\mathcal{S}[\rho^{*}]=-\frac{1}{2}\log\alpha\label{eq:Entropy_alpha}\end{equation}
 for the expression of the entropy (see calculation in Appendix B-2).
We thus conclude that the maximum-entropy solution will be the one
verifying the constraints with minimum value for $\alpha$.

\smallskip{}
We now compute $E$ and $\Gamma_{2}$ as functions of $\beta$ and
$\alpha$. For this, we compute the average vorticity using \eqref{eq:en-en-rho-crit}
in \eqref{eq:vorticite_coarse_grained}: \[
\bar{\omega}({\bf r})=\int_{-\infty}^{+\infty}\mathrm{d}\sigma\sigma\rho^{*}(\sigma,{\bf r})=\frac{\beta}{2\alpha}\bar{\psi}({\bf r}).\]
 %\sqrt{\frac{\alpha|{\mathcal{D}}|}{\pi}}\int_{-\infty}^{+\infty}\mathrm{d}\sigma\,\mathrm{e}^{-\alpha|{\mathcal{D}}|\left(\sigma-\frac{\beta\bar{\psi}({\bf r})}{2\alpha}\right)^{2}}\sigma=\sqrt{\frac{\alpha|{\mathcal{D}}|}{\pi}}\int_{-\infty}^{+\infty}\mathrm{d}\sigma\,\mathrm{e}^{-\alpha|{\mathcal{D}}|\sigma^{2}}\left(\sigma+\frac{\beta\bar{\psi}({\bf r})}{2\alpha}\right)
We thus have $\bar{\omega}({\bf r})=\Delta\bar{\psi}({\bf r})=\beta\bar{\psi}({\bf r})/(2\alpha)$,
so we can deduce that vorticity and streamfunction are proportional
to a Laplacian eigenmode: for some $n\geq1,\,\bar{\omega}({\bf r})=A\, e_{n}({\bf r})$
and $\bar{\psi}({\bf r})=-A/\lambda_{n}\, e_{n}({\bf r})$, with $\beta_{n}=-2\alpha_{n}\lambda_{n}$.
From $\mathcal{E}[\bar{\omega}]=E$, we have $E=A^{2}/(2\lambda_{n})$
and $\bar{\psi}({\bf r})=-\sqrt{2E/\lambda_{n}}\, e_{n}({\bf r})$.
From $\Gamma_{2}=\int_{-\infty}^{+\infty}\mathrm{d}\sigma\sigma^{2}\rho^{*}(\sigma,{\bf r})$
(see Appendix B-2 for detailed computation), we get $\Gamma_{2}=2\lambda_{n}E+1/(2\alpha_{n})$.
Thus, we can see that the minimum value $\alpha_{n^{*}}$ of $\alpha$
is obtained for $n^{*}=1$.

\smallskip{}
Finally, since entropy \eqref{eq:Entropy_alpha} is maximum for $\alpha\geq0$
minimum, the first eigenmode ($n=1$) is the one selected.

\smallskip{}
We are left with \[
\bar{\omega}({\bf r})=\sqrt{2\lambda_{1}E}\, e_{1}({\bf r})\quad\text{and}\quad\alpha=\frac{1}{2\left(\Gamma_{2}-2\lambda_{1}E\right)}\,,\]
so that the equilibrium entropy is \[
S_{K}\left(E,\Gamma_{2}\right)=\frac{1}{2}\log\left(\Gamma_{2}-2\lambda_{1}E\right)+\frac{\log2}{2}\,.\]
 Comparing this result with \eqref{eq:Entropie-energie-enstrophie-texte},
we can conclude that the entropy computed from the mean-field variational
problem, in the energy-enstrophy ensemble, is the same as the one
computed directly from the definition of the energy-enstrophy microcanonical
measure, through Fourier mode decomposition (finding $\bar{\omega}({\bf r})=\sqrt{2\lambda_{1}E}\, e_{1}({\bf r})$
is equivalent to finding $P_{1}(E_{1})=\delta(E-E_{1})$ and $P_{n}(E_{n})=\delta(E_{n})$).

\subsection{Ensemble inequivalence\label{sub:Ensemble-inequivalence-Energy-Enstrophie}}

From the entropy, we can compute the inverse temperature $\beta=\partial S_{K}/\partial E=-\lambda_{1}/\left(\Gamma_{2}-2\lambda_{1}E\right)\leq0$
and fugacity $\alpha=\partial S_{K}/\partial\Gamma_{2}=1/\left[2\left(\Gamma_{2}-2\lambda_{1}E\right)\right]$.
These thermodynamical coefficients are related through $\beta=-2\lambda_{1}\alpha$.
This relation shows that some couples of thermodynamical coefficients
are not obtained in the energy-enstrophy microcanonical ensemble,
in contrast to what would be expected in the thermodynamics of classical
condensed matter systems. Moreover, the determinant of the Hessian
of $S_{K}$, that is, $\partial^{2}S_{K}/\partial E^{2}\cdot\partial^{2}S_{K}/\partial\Gamma_{2}^{2}-(\partial^{2}S_{K}/\partial E\partial\Gamma_{2})^{2}$,
is zero, showing that $S_{K}$ is not strictly concave, unlike what
would be expected for an entropy in the case of short-range interacting
systems. Both these properties are signs of non-equivalence between
the microcanonical and canonical ensembles: the two ensembles would
give different predictions (see for instance \cite{Bouchet_Barre:2005_JSP,EllisHavenTurkington:2000_Inequivalence,Dauxois_Ruffo_Arimondo_Wilkens_2002LNP...602....1D}).
This case of ensemble inequivalence, for the energy-enstrophy ensembles,
is actually a case of partial equivalence (see \cite{EllisHavenTurkington:2000_Inequivalence}
for a definition).

A detailed discussion of ensemble inequivalence and related phase
transitions, for statistical equilibrium with linear relation between
vorticity and streamfunctions, including the case of the energy-enstrophy
ensemble just discussed can be found in \cite{Venaille_Bouchet_PRL_2009}.

\subsection{Comments on the Kraichnan energy-enstrophy theory\label{sub:Kraichnan}}

The term `condensation' was proposed by Kraichnan from the analysis
of the energy-enstrophy canonical ensembles \cite{Kraichnan_Motgommery_1980_Reports_Progress_Physics}.
As explained in section \ref{sub:Hamiltonian-structure},
canonical measures are not relevant for fluid systems; they may be
useful only when yielding results equivalent to the ones from microcanonical
measures. Kraichnan noticed this and worked nonetheless with canonical
ensembles, maybe because he did not know how to perform microcanonical
computations, most likely because at that time the possibility of
ensemble inequivalence was nearly unknown %
\footnote{The first observation of ensemble inequivalence was made in the astrophysical
context \cite{LyndenBell:1968_MNRAS,Hertel_Thirring_1971_AnnPhys},
while a thorough study \cite{Dauxois_Ruffo_Arimondo_Wilkens_2002LNP...602....1D,Bouchet_Barre:2005_JSP}
and understanding of the importance of ensemble inequivalence for
two-dimensional flows \cite{Smith_ONeil_Physics_Fluids_1990_Inequivalence,Kiessling_Lebowitz_1997_PointVortex_Inequivalence_LMathPhys,Eyink_Spohn_1993_JSP....70..833E,EllisHavenTurkington:2000_Inequivalence,Venaille_Bouchet_PRL_2009}
are more recent.%
}. Unfortunately, as mentioned earlier, the energy-enstrophy ensembles
display an instance of partial ensemble inequivalence. These remarks
explain the difficulties encountered by Kraichnan when analyzing the
canonical measures, and why he wrongly concluded that a statistical
mechanics approach would work only for truncated systems. Working
in microcanonical ensembles actually allows to build invariant measures
for the real (non-truncated) 2D Euler equations. If one were interested
in truncated systems, then Kraichnan's work would remain very useful.

More importantly, when looking closely at Kraichnan's
works (see, for instance, \cite{Kraichnan_Motgommery_1980_Reports_Progress_Physics}
page 565), one sees that in the canonical ensemble, a complete condensation
of the energy on the gravest mode occurs only for specific values
of the thermodynamical parameters. For most values of the thermodynamical
parameters, an important part of the energy remains on the other modes.
Still Kraichnan argued, probably from numerical observations available
at the time and from physical insight, that these cases leading to
a condensation were the most interesting ones. The microcanonical
treatment we propose here proves that a complete condensation occurs
whatever the values of the energy and of the enstrophy, in the microcanonical
ensemble. A complete condensation is actually observed in many numerical
simulations. We thus conclude that the physical insight of Kraichnan
and his concept of condensation describes the relevant physical mechanism,
but that a treatment in the microcanonical ensemble provides a much
better understanding, and overcomes the preceding contradictions.

\smallskip{}

\paragraph*{Limitations of the energy-enstrophy approach}

There is \emph{a priori} no reason to consider only the energy and
enstrophy invariants, except for being able to solve the mathematics
easily. From the discussion of section \eqref{sub:Mean_Field}, we
know that a mean-field approach is exact for the microcanonical measures
in the case of 2D Euler equations, and thus the description of any
microcanonical measure (corresponding to any set of invariants) is
not difficult.

As has been shown in previous works (see, for instance, \cite{Bouchet_Simonnet_2008}),
when taking into account all invariants, the energy will no longer
be limited to the first mode $e_{1}$. The energy-enstrophy measure
may still be a good approximation in some cases: in the limit of small
energy, for instance, most of the energy will remain in the first
few modes. The notion of condensation will thus remain valid only
roughly speaking, at a qualitative level.

By contrast, in some cases such as that of doubly periodic domains
with aspect ratio close to but different than one (see \cite{Bouchet_Simonnet_2008}),
the notion of condensation would lead to completely wrong predictions.

\section{Invariant measures of the 2D Euler equations\label{sec:Invariant-measures-2D-Euler}}

In the previous section, we have built microcanonical
measures for the 2D Euler equations and argued that they are a special
set of Young measures. In this section, we consider the dynamics of
measures and more specifically the dynamics of Young measures. We
give a direct proof that sets of Young measures, including microcanonical
ones, are invariant measures of the 2D Euler equations. For this we
derive the evolution equation verified by the characteristic functional,
in section \ref{sub:Characterisitic_functional}. This equation describes
the evolution of all the statistics of the system: it is equivalent
to the Liouville equation and includes the hierarchy of equations
describing the evolution of the statistics of the vorticity field.

This section is completely independent from the previous
one. The proof that sets of Young measures are invariant is independent
from the building of the microcanonical measure in the previous section.
However, the construction of microcanonical measures from the Liouville
theorem, in the previous section, shows that among the set of Young
measures, microcanonical measures have a specific meaning.

\subsection{Evolution of the characteristic and cumulant-generating functionals
\label{sub:Characterisitic_functional}}

For any random variable $x$, it is customary to define the characteristic
function $f\left(l\right)=\left\langle \mathrm{e}^{ilx}\right\rangle $
and the cumulant-generating function $h\left(l\right)=\log f\left(l\right)$,
where the angle brackets denote average over the measure of the random
variable $x$. In order to describe the temporal evolution of the
statistics of the vorticity field $\omega(\mathbf{r},t)$, it will
prove very useful to use a generalization of the characteristic and
cumulant-generating functions to random fields. We consider
an ensemble of initial conditions $\left\{ \omega_{0}\left(\mathbf{r}\right)\right\} $.
Each of these initial conditions evolves according to the 2D Euler
equations, defining an ensemble of solutions of the 2D Euler equations
$\{\omega(\mathbf{r},t)\}$. We define the characteristic and cumulant-generating
functionals of the ensemble $\left\{ \omega_{0}\left(\mathbf{r}\right)\right\}$,
respectively, as \[
F_{0}[l]=\left\langle \mathrm{e}^{i\int l(\mathbf{r})\omega_{0}(\mathbf{r})\mathrm{d}\mathbf{r}}\right\rangle \,\,\,\mbox{and}\,\,\, H_{0}[l]=\log F[l],\]
 where the angle brackets denote ensemble average over realizations
of $\omega_{0}$. We define similarly the characteristic and cumulant-generating
functionals of the ensemble $\{\omega(\mathbf{r},t)\}$, respectively,
as\[
F[l,t]=\left\langle \mathrm{e}^{i\int l(\mathbf{r})\omega(\mathbf{r},t)\mathrm{d}\mathbf{r}}\right\rangle \,\,\,\mbox{and}\,\,\, H[l,t]=\log F[l,t],\]
where the angle brackets still denote ensemble average over realizations
of the initial conditions $\omega_{0}$.

We now use that each realization $\omega(\mathbf{r},t)$ is a solution
to the 2D Euler equations \eqref{eq:Euler_2D_Vorticity}, in order
to derive the evolution equation for $F$. A straightforward computation,
reproduced in Appendix D-1, leads to \begin{equation}
\frac{\partial F}{\partial t}+i\iint\mathrm{d}\mathbf{r'}\,\mathrm{d}\mathbf{r}\ {\nabla}\lambda(\mathbf{r})\cdot\mathbf{G}(\mathbf{r},\mathbf{r'})\frac{\delta^{2}F}{\delta l(\mathbf{r})\delta l(\mathbf{r'})}=0\,,\label{eq:charac_funcnal}\end{equation}
 where $\mathbf{G}$ is the Green function for the velocity: \begin{equation}
\mathbf{v}(\mathbf{r})=\int\mathrm{d}\mathbf{r'}\ \mathbf{G}(\mathbf{r},\mathbf{r}')\omega(\mathbf{r}').\label{eq:velocity-Green}\end{equation}

We note that the evolution equation for the characteristic functional
\eqref{eq:charac_funcnal} is a linear equation, as is the classical
Liouville equation.

The equation for the cumulant-generating functional is also obtained
straightforwardly (see Appendix D-2): \begin{equation}
\frac{\partial H}{\partial t}+i\iint{\nabla}l({\bf r})\cdot{\bf G}({\bf r},{\bf r'})\left(\frac{\delta^{2}H}{\delta l({\bf r})\delta l({\bf r'})}+\frac{\delta H}{\delta l({\bf r})}\frac{\delta H}{\delta l({\bf r'})}\right)\,\mathrm{d}{\bf {r'}}\,\mathrm{d}{\bf {r}}=0\,.\label{eq:Evolution_H}\end{equation}

\subsection{Young measures and their dynamics}

\subsubsection{Young measures}

We recall that Young measures are\textbf{ }uncountable\textbf{ }product
measures: the probability distribution of the vorticity field at\textbf{
}an arbitrary number of points $\left\{ \mathbf{r}_{k}\right\} $
is given by the product of the independent measures\textbf{ }$\rho(\sigma,{\bf \mathbf{r}}_{k})$
at each point $\mathbf{r}_{k}$. We note that at each point, $\rho$
is normalized\[
\int_{-\infty}^{+\infty}\mathrm{d}\sigma\,\,\rho\left(\sigma,\mathbf{r}\right)=1.\]
As we see below, the fact that vorticity values at different points
are independent variables has important consequences.

The set of deterministic vorticity fields are $\omega\left(\mathbf{r}\right)$
is a special class of Young measures with $\rho\left(\sigma,\mathbf{r}\right)=\delta\left(\sigma-\omega\left(\mathbf{r}\right)\right)$.

The set of microcanonical measures described in section \textbf{\eqref{sub:Mean_Field},}
is a special class of Young measures. They are defined as \begin{equation}
\rho_{\beta,\left\{ \alpha\right\} }(\sigma,{\bf r})=\frac{1}{Z(\beta\psi\left(\mathbf{r}\right))}\mathrm{e}^{\beta\sigma\psi\left(\mathbf{r}\right)-\alpha(\sigma)},\label{eq:microcanonical}\end{equation}
 where $Z_{\alpha}(u)=\int_{-\infty}^{+\infty}\mathrm{d}\sigma\ \mathrm{e}^{\sigma u-\alpha(\sigma)}$.

If the local probabilities $\rho_{1}(\sigma,{\bf \mathbf{r}})$ and
$\rho_{2}(\sigma,{\bf \mathbf{r}})$ define two Young measures, then
for any $0\leq l\leq1$, $l\rho_{1}+(1-l)\rho_{2}$ defines a Young
measure.

\subsubsection{Cumulant-generating functionals for Young measures\label{sub:Cumulant-gen-Young}}

Let us evaluate the cumulant-generating functional of a Young measure.
For this purpose, we consider $h\left(l,\mathbf{r}\right)$, the cumulant-generating
function of the local probability $\rho(\sigma,{\bf r})$ at each
point $\mathbf{r}$: \begin{equation}
h\left(l,\mathbf{r}\right)=\log f\left(l,\mathbf{r}\right)\,\,\,\mbox{with}\,\,\, f\left(l,\mathbf{r}\right)=\int_{-\infty}^{+\infty}\mathrm{d}\sigma\ \mathrm{e}^{il\sigma}\rho(\sigma,\mathbf{r}),\label{eq:h}\end{equation}
 and the average vorticity field \begin{equation}
\bar{\omega}(\mathbf{r})=\int_{-\infty}^{+\infty}\mathrm{d}\sigma\ \sigma\rho(\sigma,\mathbf{r})=\frac{\partial h}{\partial l}\left(0,\mathbf{r}\right).\label{eq:Omega_Moyen}\end{equation}

Let us now consider a sufficiently regular $\lambda(\mathbf{r})$
(for instance, Riemann-integrable for the purpose of the following
discussion). The characteristic functional of the entire vorticity
field is easily computed using an approximation by a finite
Riemann sum: \[
F[\lambda]=\left\langle \mathrm{e}^{i\int\lambda(\mathbf{r})\omega(\mathbf{r})\mathrm{d}\mathbf{r}}\right\rangle =\lim_{N\to\infty}\left\langle \mathrm{e}^{\frac{i|{\mathcal{D}}|}{N}\sum_{k=1}^{N}\lambda(\mathbf{r}_{k})\omega(\mathbf{r}_{k})}\right\rangle,\]
where $|\mathcal{D}|$ still denotes the domain area. Since variables
$\{\omega(\mathbf{r}_{k})\}_{k}$ are independent, we use the fact
that the characteristic function of a set of independent variables
is equal to the product of the single-variable characteristic functions:
\[
F[\lambda]=\lim_{N\to\infty}\prod_{k=1}^{N}f\left(\frac{|{\mathcal{D}}|\lambda\left(\mathbf{r}_{k}\right)}{N},\mathbf{r}_{k}\right),\]
 so that \[
H[\lambda]=\log F[\lambda]=\lim_{N\to\infty}\sum_{k=1}^{N}h\left(\frac{i|{\mathcal{D}}|}{N}\lambda(\mathbf{r}_{k}),\mathbf{r}_{k}\right)=\lim_{N\to\infty}\frac{i|{\mathcal{D}}|}{N}\sum_{k=1}^{N}\frac{\partial h}{\partial\lambda}\left(0,\mathbf{r}_{k}\right)\lambda(\mathbf{r}_{k})\,,\]
 where we have used $h\left(0,\mathbf{r}_{k}\right)=0$. Then, using
(\ref{eq:Omega_Moyen}), we have
\begin{equation}
H[\lambda]=i\int\lambda(\mathbf{r})\bar{\omega}(\mathbf{r})\mathrm{d}\mathbf{r}.\label{eq:Cumulant-generating-functionnal-smooth}
\end{equation}
The cumulant-generating functional is linear in $\lambda$. Actually,
comparing \eqref{eq:Cumulant-generating-functionnal-smooth} to \eqref{eq:H-deterministe},
we see that for sufficiently regular $\lambda\left(\mathbf{r}\right)$,
the cumulant-generating functional is the same as that of a deterministic
field with vorticity the average vorticity $\bar{\omega}(\mathbf{r})$.
Hence, the statistics of any observable that is obtained as\textbf{
}the domain integral of a sufficiently smooth function of \textbf{$\omega(\mathbf{r})$}
depends only on the average vorticity $\bar{\omega}(\mathbf{r})$.
This can be seen as an example of a law of large numbers for an infinite
sum of independent variables.

In the following, we call an observable $S$ smooth additive, if $S$
is obtained as an integral over the vorticity field\textbf{:} \begin{equation}
S=\int\mathrm{d}\mathbf{r}\,\phi(\mathbf{r})\omega(\mathbf{r}).\label{eq:S}\end{equation}
 Then, from the preceding result, we can conclude that the distribution
of any smooth additive observable $S$ is a delta function. As an
illustration, we compute the characteristic functional of the velocity
field at point $\mathbf{r}$, using $\mathbf{\lambda}(\mathbf{r}')=\lambda\mathbf{G}\left(\mathbf{r},\mathbf{r}'\right)$,
where $\mathbf{G}$ is the velocity Green function (\ref{eq:velocity-Green}).
Then, \begin{equation}
\log\left\langle \mathrm{e^{i\lambda\mathbf{v}\left(\mathbf{r}\right)}}\right\rangle =H[\lambda\mathbf{G}\left(\mathbf{r},\mathbf{r}'\right)]=i\lambda\int\mathrm{d}\mathbf{'r}\,\mathbf{G}\left(\mathbf{r},\mathbf{r}'\right)\bar{\omega}(\mathbf{r'})\equiv i\lambda\bar{\mathbf{v}}\left(\mathbf{r}\right).\label{eq:average_velocity}\end{equation}
We see that the cumulant-generating function of $\mathbf{v}\left(\mathbf{r}\right)$
is linear in $\lambda$. The velocity field has no fluctuations, and
hence is a delta function centered at the average value $\bar{\mathbf{v}}\left(\mathbf{r}\right)$.
Once again, this is nothing but the law of large numbers.\\

We can easily generalize the computation of the cumulant-generating
functional (\ref{eq:Cumulant-generating-functionnal-smooth}) to classes
of non-regular $\lambda(\mathbf{r})$. As an example, we consider
$\lambda\left(\mathbf{r}\right)=L\left(\mathbf{r}\right)+l\delta\left(\mathbf{r}-\mathbf{r}_{0}\right)$,
where $L\left(\mathbf{r}\right)$ is Riemann-integrable and $l$ is
a scalar. A direct generalization of the computations preceding (\ref{eq:Cumulant-generating-functionnal-smooth})
leads to \[
H[L(\mathbf{r})+l\delta(\mathbf{r}-\mathbf{r}_{0})]=i\int L(\mathbf{r})\bar{\omega}(\mathbf{r})\mathrm{d}\mathbf{r}+h\left(l,\mathbf{r}_{0}\right).\]
Using this result with $L(\mathbf{r})=\lambda\phi$, we can describe
the joint probability of any smooth additive variable $S$ (\ref{eq:S})
and of the vorticity at point $\mathbf{r}_{0}$. We conclude that
$S$ and $\omega\left(\mathbf{r}_{0}\right)$ are independent random
variables, $\omega\left(\mathbf{r}_{0}\right)$ having distribution
$\rho\left(\sigma,{\bf r}\right)$ and $S$ having a delta distribution
centered at the average value $\bar{S}=\int\mathrm{d}\mathbf{r}\,\phi(\mathbf{r})\bar{\omega}(\mathbf{r})$.

This result can be extended to the joint probability distribution
of the vorticity at an arbitrary number of points and of any smooth
additive observable. For instance, we can conclude that for Young
measures, the velocity field is a random variable independent of the
vorticity field, having a delta distribution centered at the average
velocity $\bar{\mathbf{v}}(\mathbf{r})$, with $\Delta\bar{\psi}=\bar{\omega}$.

\subsubsection{Dynamics of Young measures\label{sub:Dynamics-of-Young-Measures}}

We obtain the dynamics of $h(\lambda,\mathbf{r})$, the cumulant-generating
function of the vorticity field at point $\mathbf{r}$ (\ref{eq:h}),
either from (\ref{eq:Evolution_H}) or by direct averaging of the
2D Euler equations. Using that the velocity field is independent of
the vorticity field, as explained at the end of section \ref{sub:Cumulant-gen-Young},
we obtain \begin{equation}
\frac{\partial h}{\partial t}+\bar{\mathbf{v}}\cdot\nabla h=0\,.\label{eq:h-transport-young}\end{equation}
From this equation, it is clear that any initial measure which is
Young measure remains a Young measure over time. Moreover equation
(\ref{eq:h-transport-young}) describe the whole dynamics, as for
Young measures, any observable can be derived from $h$. The dynamics
of Young measures is thus rather simple.

For the dynamics of Young measures, we could have worked directly
with $\rho$. The evolution equation for $\rho$ is just \[
\frac{\partial\rho}{\partial t}+\bar{\mathbf{v}}\cdot\nabla\rho=0\,.\]
 However it is more convenient to work with $H$ (and $h$) as soon
as perturbations to a Young measure are considered.

\subsection{Classes of invariant measures}

\subsubsection{Deterministic dynamical equilibria}

For any dynamical system, equilibria of the deterministic equations
are trivial invariant measures. This is obviously also the case for
the 2D Euler equations.

Let us consider a stationary solution of the 2D Euler equations $\omega_{0}(\mathbf{r})$,
and the associated velocity field $\mathbf{v}_{0}({\bf r})$, such
that $\mathbf{v}_{0}({\bf r}) \cdot \nabla\omega_{0}(\mathbf{r})=0.$
The associated characteristic and cumulant-generating functionals
are $F_{0}[\lambda]=\mathrm{e}^{i\int\mathrm{d}\mathbf{r}\,\lambda(\mathbf{r})\omega_{0}(\mathbf{r})}$
and \begin{equation}
H_{0}[\lambda]=i\int\mathrm{d}\mathbf{r}\,\lambda(\mathbf{r})\omega_{0}(\mathbf{r})\label{eq:H-deterministe}\end{equation}
 respectively. It is easily verified that $F_{0}$ and $H_{0}$ are
equilibria of (\ref{eq:charac_funcnal}) and (\ref{eq:Evolution_H}),
respectively, as expected. Indeed, the second term in the l.h.s. of
equation (\ref{eq:Evolution_H}) is then \[
-i\int\mathrm{d}\mathbf{r}\,\nabla\lambda(\mathbf{r})\cdot\mathbf{v}_{0}({\bf r})\omega_{0}(\mathbf{r}),\]
 which is trivially null. This is also trivially checked from (\ref{eq:h-transport-young}).

\subsubsection{Invariant Young measures}

From (\ref{eq:Evolution_H}) and the cumulant-generating functional
(\ref{eq:Cumulant-generating-functionnal-smooth}), we see that a
necessary condition for a Young measure to be invariant for the 2D
Euler equations is that for any sufficiently regular $\lambda(\mathbf{r})$,
\[
\int\mathrm{d}\mathbf{r}\,\nabla\lambda(\mathbf{r})\cdot\bar{\mathbf{v}}({\bf r})\bar{\omega}(\mathbf{r})=0\,.\]
 Thus, a necessary condition is that the average vorticity $\bar{\omega}(\mathbf{r})$
be a dynamical equilibrium of the 2D Euler equations. From this equation,
we see that $h$ is transported by the average velocity $\bar{\mathbf{v}}$.
Then a further necessary condition for a Young measure to be invariant
is that $h$ be invariant over any streamline of the velocity $\bar{\mathbf{v}}$.

This is also a sufficient condition. Indeed, in 2D Euler equations,
vorticity is just transported by the velocity field. Then, because
for a Young measure, the velocity has no fluctuations, if the velocity
is moreover stationary and if the distribution does not depend on
the streamline, then the Young measure is invariant. Then, invariant
Young measures are the ones for which $h$ is invariant over streamlines
of $\bar{\mathbf{v}}$.

A smaller class of invariant Young measures of interest is the one
for which $\rho$ depends in a functional way on the streamfunction
$\psi=\bar{\psi}\,:\;\rho=\rho(\sigma,\bar{\psi}\left(\mathbf{r}\right))$.
This property has to be self-consistent: \[
\bar{\omega}=\Delta\bar{\psi}=\int d\sigma\,\sigma\rho(\sigma,\bar{\psi}\left(\mathbf{r}\right)).\]

\subsubsection{Microcanonical measures}

The set of microcanonical measures (\ref{eq:microcanonical}) is a
special class of Young measures. From the previous computations, their
cumulant-generating functional is

\[
H[\lambda]=i\int\lambda(\mathbf{r})\bar{\omega}(\mathbf{r})\mathrm{d}\mathbf{r}\quad\text{with}\,\,\,\bar{\omega}(\mathbf{r})=g\left(\bar{\psi}\left(\mathbf{r}\right)\right)\,\,\,\text{and}\,\,\, g\left(u\right)=\frac{\mathrm{d}}{\mathrm{d}u}\log Z\,.\]
 Because of the functional relation between vorticity and streamfunction
$\bar{\omega}(\mathbf{r})=g\left(\bar{\psi}\left(\mathbf{r}\right)\right)$,
$\bar{\omega}(\mathbf{r})$ is a dynamical equilibrium of the 2D Euler
equations. In conclusion, the microcanonical measures are invariant
measures of the 2D Euler equations.

\subsubsection{Quasi-invariant Young measures\label{par:Quasi-invariant-Young-measures}}

We now consider the class of Young measures for which $\bar{\omega}(\mathbf{r})$
is a dynamical invariant of the 2D Euler equations, but for which
$h(\lambda,\mathbf{r})$ is not invariant over each streamline. Then,
from (\ref{eq:h-transport-young}), because velocity $\bar{\mathbf{v}}(\mathbf{r})$
is stationary, $h(\lambda,\mathbf{r})$ is just transported along
each streamline. Therefore, from a microscopic point of view, such
a Young measure is not invariant, but from a macroscopic point of
view, it is: any smooth additive observable of the vorticity field,
including the velocity field, is invariant. We call such a measure
a quasi-invariant Young measure.

\subsection{Ergodicity}

Section \ref{sec:Equilibrium-statistical-mechanics} describes the
statistical equilibria through the variational problem (\ref{eq:MVP}).
The solution of this variational problem is the most probable state
and also, thanks to the large-deviation property, the state around
which an overwhelming majority of states do concentrate, for the microcanonical
measure. Besides, the microcanonical measure is the most natural invariant
measure of the 2D Euler equations with the dynamical constraints.

Having described a natural invariant measure of the equations is an
important theoretical step. Another important point would be to know
if this invariant measure is the only one having the right values
for the dynamical invariants. The evolution of one trajectory of the
dynamical system also defines a measure (through time averaging).
If we knew the invariant measure were unique, then it would mean that
averaging over the microcanonical measure is equivalent to averaging
over time. When this uniqueness property holds, we call the dynamical
system ergodic.

Generally speaking, the ergodicity of a dynamical system is a property
that is usually extremely difficult to prove. Such proofs exist only
for very few extremely simple systems. Ergodicity is actually thought
to be wrong in general. For instance, in Hamiltonian systems with
a finite number of degrees of freedom, there often exist islands in
phase space in which trajectories are trapped. The common belief in
the statistical mechanics community is that those parts of phase space
where the motion is trapped exist, but occupy an extremely small relative
volume of the phase space, for generic systems with a large number
of degrees of freedom. Apart from a few systems which were proved
to be integrable, this common wisdom has successfully passed empirical
tests of a century of statistical mechanics studies.\\

There is no reason to suspect that this general picture should be
different in the case of the 2D Euler equations, in general. It is
thus thought that an overwhelming number of initial conditions will
have a dynamics consistent with the microcanonical measure predictions.
However, similarly to most other Hamiltonian systems, the 2D Euler
equations are actually non-ergodic, the proofs being extremely simple,
given the discussion on invariant Young measures in the previous section.

Indeed, in section \ref{sub:Characterisitic_functional},
we have proved that any Young measure for which $\bar{\omega}\left(\mathbf{r}\right)$
is a stationary solution of the 2D Euler equations is either an invariant
or a quasi-invariant measure. The class of invariant measures corresponding
to ensemble of trajectories with given values of the invariants, is
then much, much larger than the class of statistical equilibrium invariant
measures with the same invariants. This proves that nontrivial sets
of vorticity fields are dynamically invariant. In this restricted
sense, this proves that the 2D Euler equations are not ergodic.

This theoretical argument proving non-ergodicity is in accordance
with previous remarks about the phenomenology of the 2D Euler or quasi-geostrophic
equations. For instance, it was observed numerically that initial
conditions with localized vorticity, in large domains, remain localized
(see \cite{Chavanis_Sommeria_1998JFM_LocalizedEquilibria...356..259C}
and references therein; \cite{Chavanis_Sommeria_1998JFM_LocalizedEquilibria...356..259C}
actually proposes an interesting phenomenological modification of
the microcanonical measure approach to cope with this localized dynamics
problem). Another example of possible non-ergodicity is the dynamics
close to stable dynamical equilibria of the equations. When trajectories
come close to such equilibria, they can be trapped (frozen) as was
seen in some numerical simulations. A classical argument by Isichenko
\cite{Isichenko_1997_PhRvL} is that for initial conditions close
to parallel flows, `displacement in certain directions is uniformly
small, implying that decaying Vlasov and 2D fluid turbulence are not
ergodic'. Even if the predicted algebraic laws by Isichenko are
most probably wrong, the fact that displacement in directions normal
to the streamlines is uniformly small is probably right, thus being
another argument for non-ergodicity.\\

An important point to be noted, is that the Navier-Stokes equation
with stochastic forces can be proved to be ergodic \cite{Bricmont_Kupianen_2001_Comm_Math_Phys_Ergodicity2DNavierStokes}.
This ergodicity refers to invariant measures of the Navier-Stokes
equations, which are non-equilibrium invariant measures with fluxes
of conserved quantity. A very important point is to understand the
limit of weak forces and dissipation for such invariant measures and
to study their relations with the invariant measures of the 2D Euler
equations. Some very interesting results can be found in \cite{Kuksin_2004_JStatPhys_EulerianLimit}.

\subsection{Stability of invariant measures}

As discussed in the previous section, the dynamical stability of invariant
measures is an essential point. In the following two sections, we
give a first discussion of the dynamical stability of invariant Young
measures.

We first discuss, in section \ref{sub:Stability-among-Young-measures},
the stability of invariant Young measures when the perturbation is
such that the initial condition is still a Young measure. We conclude
that the stability then depends only on the unperturbed average vorticity
$\bar{\omega}_{0}\left(\mathbf{r}\right)$ and velocity $\bar{\mathbf{v}}_{0}(\mathbf{r})$.
More precisely, if in the framework of the 2D Euler equations small
perturbations of $\bar{\omega}\left(\mathbf{r}\right)$ lead to finite
Lagrangian transport in the direction transverse to the streamlines,
then the Young measure is stable to perturbations among Young measures.
This condition of finite Lagrangian transport cross to the streamlines,
is true for the 2D linearized equations for a whole class of parallel
flows, including flows with stationary streamlines \cite{Bouchet_Morita_2010PhyD}.
Moreover, we guess it is also true for the nonlinear 2D Euler equations,
for large classes of parallel flows and stable vortices, even if there
are still no proofs yet.

In section \ref{sub:Stability-for-small-velocity}, we consider the
stability of dynamical equilibria of the 2D Euler equations, subjected
to perturbations whose measure is not necessarily in the the class
of Young measures. For this study, we assume perturbations to be small.
We prove that the linear stability of the unperturbed flow in the
framework of the 2D Euler equations, implies the linear stability
in a statistical sense.

\subsubsection{Stability among Young measures\label{sub:Stability-among-Young-measures}}

As explained in section \ref{sub:Dynamics-of-Young-Measures}, the
dynamics of a Young measure is equivalent to the dynamics of its cumulant-generating
function $h(\lambda,\mathbf{r})$ (\ref{eq:h-transport-young}), where
$\bar{\mathbf{v}}(\mathbf{r})$ is the velocity field corresponding
to $\bar{\omega}(\mathbf{r})$. Moreover, the set of Young measures
is dynamically stable. We can thus consider stability of Young measures
among the set of Young measures (we take the initial condition as
a Young measure that is close to the unperturbed invariant one $h_{0}(\lambda,\mathbf{r})$).
We note that this includes the case when each realization leading
to the Young measure is perturbed by the same initial perturbation
$\delta\omega\left(\mathbf{r}\right)$, in which case $h(\lambda,\mathbf{r})=h_{0}(\lambda,\mathbf{r})+i\lambda\delta\omega\left(\mathbf{r}\right)$.

We have seen that invariant Young measures are the ones for which
$\bar{\mathbf{v}}(\mathbf{r})$ is stationary and $h(\lambda,\mathbf{r})$
is constant over every streamline of $\bar{\mathbf{v}}(\mathbf{r})$.
We consider such an invariant Young measure, denoting the associated
cumulant-generating function by $h_{\text{0}}(\lambda,\mathbf{r})$
and the associated velocity by $\bar{\mathbf{v}}_{0}(\mathbf{r})$,
and a small perturbation to $h_{0}:\, h=h_{\text{0}}+\delta h$. We
recall (see (\ref{eq:h-transport-young})), that the dynamics is
\begin{equation}
\frac{\partial h}{\partial t}+\bar{\mathbf{v}}\cdot\nabla h=0,\label{eq:Dynamics_Young_2}
\end{equation}
with $\bar{\omega}(\mathbf{r})=\frac{\partial h}{\partial\lambda}\left(0,\mathbf{r}\right)$
and $\bar{\mathbf{v}}(\mathbf{r})$ the corresponding velocity field.
From (\ref{eq:Dynamics_Young_2}), we have
\begin{equation}
\frac{\partial\bar{\omega}}{\partial t}+\bar{\mathbf{v}}\cdot\nabla\bar{\omega}=0.\label{eq:omega-moyen-transport-young}\end{equation}
 We conclude that $\bar{\omega}$ satisfies the deterministic 2D Euler
equations. A necessary condition for $h_{0}$ to be stable is thus
that $\omega_{\text{0}}$ be stable for the 2D Euler equations.

A variety of notions of stability exist for the 2D Euler equations,
depending on the norms used to control the initial conditions and
the evolving\textbf{ }solutions. As far as the\textbf{ }initial conditions
are concerned, from a physical point of view, it is very natural to
consider small perturbations of the initial velocity, as small-scale
vorticity fluctuations are usually not controlled. We note that (\ref{eq:Dynamics_Young_2})
and (\ref{eq:omega-moyen-transport-young}) are readily solved using
Lagrangian coordinates: defining $\mathbf{R}(t)$ by \[
\frac{\mathrm{d}\mathbf{R}}{\mathrm{d}t}=\mathbf{v}(\mathbf{\mathbf{R}})\quad\text{with\;}\mathbf{R}(t=0)\mathbf{r}=\mathbf{r}.\]

Then, using the incompressibility of $\mathbf{v}$, we deduce that$\mathbf{R}(t)$
is invertible at each time $t$, and we have: $\omega(\mathbf{r},t)=\omega(\mathbf{R}(t)^{-1}\mathbf{r},0)$
and $h(\lambda,\mathbf{r},t)=h(\lambda,\mathbf{R}(t)^{-1}\mathbf{r},0)$.

As discussed in paragraph \ref{par:Quasi-invariant-Young-measures},
the motion along streamlines does not matter for the invariance of
the measure. A natural definition of stability is therefore\textbf{
}imposing that the motion normal to the streamlines remain small over
time. We call this Lagrangian stability; it can be defined more precisely
as, say,

\[
\forall\,\epsilon,\;\exists\,\alpha:\quad\frac{1}{2}\int_{\mathcal{D}}\ \mathrm{d}\mathbf{r\,}(\delta\mathbf{v})^{2}(t=0)\leq\alpha\quad\Rightarrow\quad\lVert\nabla_{n}^{\perp}\omega_{0}\cdot\mathbf{R}(t)\mathbf{r}\lVert\leq\epsilon,\]

where $\nabla_{n}^{\perp}\omega_{0}$ is the unit vector orthogonal
to $\nabla\omega_{0}$ and where different norms $\lVert.\lVert$
define different notions of stability.

Then, clearly, from the preceding discussion, because the dynamics
of Young measures is just the transport by the average velocity, and
$h(\lambda,\mathbf{r},t)=h(\lambda,\mathbf{R}(t)^{-1}\mathbf{r},0)$,
it is natural to define the Lagrangian stability of the Young measure,
just as to be equivalent to the Lagrangian stability of the of 2D
Euler equations for the average vorticity field. \\

As can be readily seen from the results in \cite{Bouchet_Morita_2010PhyD},
a whole class of parallel flows, including flows with non-monotonic
velocity profiles, are Lagrangian-stable, as far as the linear dynamics
is concerned (if the perturbed velocity field evolves according to
the linearized 2D Euler equations, then the associated Lagrangian
transport in the direction transverse to the streamlines is uniformly
bounded over time and proportional to the initial perturbation amplitude).
Even if there is no proof yet, we guess that this is also true for
the (nonlinear) 2D Euler dynamics for a large class of parallel flows
or stable vortices. This would prove the stability of Young measures
for the nonlinear dynamics.

Then, any further discussion needs a detailed study of the relaxation
(asymptotic stability) of the 2D Euler equations, which is not available
yet but will be considered in future works.

\subsubsection{Stability of invariant measures to small velocity perturbations\label{sub:Stability-for-small-velocity}}

We consider now the stability of a deterministic solution to the 2D
Euler equations and the effect of small perturbations. The statistics
is not limited to Young measures.

We start with the 2D Euler equation which is verified for any realization:
\[
\frac{\partial\Omega}{\partial t}+\mathbf{V}\cdot\mathbf{\nabla}\Omega = 0,
\]
 where $\Omega=\Omega_{0}+\varepsilon\omega\,,\,{\bf V}={\bf V_{0}}+\varepsilon{\bf v}$,
and ${\bf V_{0}}\cdot\nabla\Omega_{0}=0$ ($\varepsilon$ is the perturbation
amplitude). Developing these equations yields
\begin{equation}
\frac{\partial\omega}{\partial t}+L[\omega]+\varepsilon{\bf v}\cdot\nabla\omega=0,\label{eq:linearized-transport}
\end{equation}
where the operator $L$, giving the linearized 2D Euler equations, is defined %publisher!!
by \begin{equation}
L[\omega]={\bf V_{0}}\cdot\nabla\omega+{\bf v}\cdot\nabla\Omega_{0}={\bf V_{0}}\cdot\nabla\omega+\left(\int\mathrm{d}{\bf {r'}}\ {\bf G}({\bf r},{\bf r'})\omega({\bf r'})\right)\cdot\nabla\Omega_{0}.\label{eq:operateur_Ldroit}
\end{equation}
The evolution equation for the characteristic functional $F[\lambda]=\langle\mathrm{e}^{i\int\lambda(\mathbf{r})\omega(\mathbf{r})\mathrm{d}\mathbf{r}}\rangle$
is then

\begin{equation}
\frac{\partial F}{\partial t}+\iint\mathrm{d}{\bf {r'}}\,\mathrm{d}{\bf {r}}\ {\nabla}\lambda({\bf r})\cdot\left[{\bf G}({\bf r},{\bf r'})\left(\varepsilon i\frac{\delta^{2}F}{\delta\lambda({\bf r})\delta\lambda({\bf r'})}-\Omega_{0}({\bf r})\frac{\delta F}{\delta\lambda({\bf r'})}\right)-{\bf V_{0}}({\bf r})\frac{\delta F}{\delta\lambda({\bf r})}\right] = 0.\label{eq:charac_funcnal_near-eq}\end{equation}
(a detailed derivation is provided in Appendix D). Likewise, the
cumulant-generating functional $H=\ln F$ satisfies
\begin{align}
\frac{\partial H}{\partial t}+\iint\mathrm{d}{\bf {r'}}\,\mathrm{d}{\bf {r}}\ {\nabla}\lambda({\bf r})\cdot\left[{\bf G}({\bf r},{\bf r'})\left(\varepsilon i\left(\frac{\delta^{2}H}{\delta\lambda({\bf r})\delta\lambda({\bf r'})}+\frac{\delta H}{\delta\lambda({\bf r})}\frac{\delta H}{\delta\lambda({\bf r'})}\right)-\Omega_{0}({\bf r})\frac{\delta H}{\delta\lambda({\bf r'})}\right)-{\bf V_{0}}({\bf r})\frac{\delta H}{\delta\lambda({\bf r})}\right] = 0.\label{eq:mom-gen_funcnal_near-eq}\end{align}
\smallskip\par
We expand the equation for the cumulant-generating functional in
powers of $\epsilon$: $H=H_{0}+\varepsilon H_{1}+\varepsilon^{2}H_{2}+\ldots$.
At lowest order ($\varepsilon^{0}$) we have
\begin{equation}
\frac{\partial H_{0}}{\partial t}+\mathcal{L}\!\left[\frac{\delta H_{0}}{\delta\lambda({\bf r})}\right]=0,\label{eq:pert_exp_lowest-order}
\end{equation}
where the linear operator $\mathcal{L}$ is defined by \begin{align}
\mathcal{L}\!\left[\frac{\delta H}{\delta\lambda({\bf r})}\right] & =-\int\mathrm{d}{\bf r}\ \nabla\lambda({\bf r})\cdot\left[{\bf V_{0}}({\bf r})\frac{\delta H}{\delta\lambda({\bf r})}+\Omega_{0}({\bf r})\int\mathrm{d}{\bf r'}\ {\bf G}({\bf r},{\bf r'})\frac{\delta H}{\delta\lambda({\bf r'})}\right]\nonumber \\
 & =\int\mathrm{d}{\bf r}\ \lambda({\bf r})L\!\left[\frac{\delta H}{\delta\lambda({\bf r})}\right]\end{align}
(see equation \eqref{eq:operateur_Ldroit}). We now remark that $H_{0}$
satisfies the same equation as the one for the cumulant-generating
functional in the case of linearized 2D Euler equations.

At the linear level (first order in $\epsilon$), the stability of
the measure is thus equivalent to the stability of the operator $\mathcal{L}$.
We remark that a detailed knowledge of the properties of \textbf{$L$}
is sufficient to describe the properties of $\mathcal{L}$; indeed,\[
\exp\left(t\mathcal{L}\right)\!\left[\frac{\delta H}{\delta\lambda({\bf r})}\right]=\int\mathrm{d}{\bf r}\ \lambda({\bf r})\exp\left(tL\right)\!\left[\frac{\delta H}{\delta\lambda({\bf r})}\right].\]
As a consequence, the stability of $L$ ($\exp\left(tL\right)$ uniformly
bounded over time) is a necessary and sufficient condition for the
stability of $\mathcal{L}$ ($\exp\left(t\mathcal{L}\right)$ uniformly
bounded over time). In the case of stable parallel base flow ${\bf V_{0}}=U\left(y\right)\mathbf{e}_{x},$
a detailed study of the asymptotic behavior of the linear operator
$L$ is provided in \cite{Bouchet_Morita_2010PhyD}: in particular, for
any perturbation, it is proved that $\lVert\exp\left(tL\right)\lVert$
is usually composed of the contribution of few modes (often no mode,
actually) plus the contribution of a continuous spectrum that decays
algebraically for large times. These results hold in the cases of
both monotonic and non-monotonic velocity profiles $U$, and are probably %publisher!!
easily generalizable to the case of stable circular vortices.

We note that the expansion of $H$ at higher orders could be performed
easily. Discussion of the convergence of such an expansion requires
a detailed study of the relaxation (asymptotic stability) of the (nonlinear)
2D Euler equations, which is not available yet and that will be considered
in future works.\\

As already stated several times, we note that a complete theory for
the stability of Young measures requires the understanding of the
nonlinear relaxation of the 2D Euler equations. Even at a linear level,
a more complete study of the stability of Young measures would involve
investigating the effect of small perturbations on any invariant Young
measure, not only on deterministic solutions as done in this section.
This requires a more involved expansion than the one in this section
and will also be considered later on.

\section{Invariant measures of the Vlasov equation\label{sec:Invariant-measures-Vlasov}}

In this section, we consider the Vlasov equation. For the sake of
simplicity we consider one-dimensional physical systems: however, the
discussion easily extends to any dimension. Thanks to the the theoretical
similarity between the 2D Euler and the Vlasov equations, noted decades
ago, all the discussions about the microcanonical measures, Young
measures, invariant measures, and stability of invariant measures
of sections \ref{sec:Equilibrium-statistical-mechanics}, \ref{sub:Direct_Comp_energie-enstrophie},
and \ref{sub:Characterisitic_functional} easily extend to the case
of the Vlasov equation. In the following sections, we only describe
briefly the statistical equilibrium measures and the invariant Young
measures. We also give a proof of the uniqueness of statistical equilibria
in the case of repulsive convex potentials, which can be useful in
many future studies.

\subsection{The Vlasov equation}

We consider a set of particles subjected to their mutual two-body
interactions with potential $W$. Each particle located at point $x$
is subject to the potential $\phi_{\mathrm{discrete}}(x)=\frac{1}{N}\sum_{i=1}^{N}W(x-x_{i})$,
where $\left\{ x_{i}\right\} $ are the particle positions. $W$ is
an even function. Classical physical arguments and mathematical proofs
justify that, when $W$ is regular enough, it is natural to consider
the following continuum approximation to this potential:
\begin{equation}
\phi(x,t)=W(x-x')f(x',p,t).\label{eq:phi}
\end{equation}
The time evolution for the one-particle phase space distribution
function $f(x,p,t)$ satisfies the Vlasov equation, given by \begin{equation}
\frac{\partial f}{\partial t}+p\frac{\partial f}{\partial x}-\frac{\mathrm{d}\phi}{\mathrm{d}x}\frac{\partial f}{\partial p}=0\,.\label{eq:vlasov}\end{equation}
If $\left\{ x_{i}\right\} $ were $N$ independent random variables
distributed according to the distribution $f$, equation \eqref{eq:phi}
would then follow from the law of large numbers and would be a good
approximation to $\phi_{\mathrm{discrete}}$ up to corrections of
order $1/\sqrt{N}$. Replacing the true discrete potential by $\phi$
thus amounts to neglecting correlations between particles (the equivalent
of the Stosszahl Ansatz) and finite-$N$ effects. The potential $\phi_{\mathrm{discrete}}$
being replaced by an average one, namely $\phi$, may be seen as a
mean-field approximation to the dynamics.

As can be easily verified, the Vlasov equation \eqref{eq:vlasov}
inherits the conservation laws of the microscopic Hamiltonian dynamics,
for instance, for the energy \begin{equation}
H[f]=\int\mathrm{d}x\mathrm{d}p\left[f\frac{p^{2}}{2}+\frac{f\phi\left[f\right]}{2}\right],\label{eq:Energie_Continue}\end{equation}
 and for the linear or angular momentum, when the system has the corresponding
translational or rotational symmetry, respectively.

If we define $\psi(x,p)=-(p^{2}/2+\phi(x))$ and \[
{\bf v}=\left(-\frac{\partial\psi}{\partial p},\frac{\partial\psi}{\partial x}\right)\quad\text{and}\quad{\bf \nabla}=\left(\frac{\partial}{\partial x},\frac{\partial}{\partial p}\right),\]
 then, the Vlasov equation \eqref{eq:vlasov} can be recast into \begin{equation}
\frac{\partial f}{\partial t}+{\mathbf{v}}\cdot{\mathbf{\nabla}}f=0\,,\label{eq:Euler-Vlasov}\end{equation}
 with ${\bf \nabla}\cdot{\bf v}=0$\textbf{ }(see equations \eqref{eq:Euler_2D_Vorticity}).
This simple remark explains the deep analogy between the 2D Euler
and the Vlasov equations. Like the vorticity for 2D flows, $f$ is
transported by an incompressible flow. This explains most of the following
properties.\\

The Casimir functionals \begin{equation}
C_{s}[f]=\int\mathrm{d}x\mathrm{d}p~s\left(f(x,p,t)\right)\label{eq:casimirs}\end{equation}
 are invariant \emph{for any function} $s$.

Let $f_{m}=\sup\{f\}$. We denote $\bar{A}(\sigma)$ the area of $\mathcal{D}$
with $f$ values greater than $\sigma$, and $\gamma\left(\sigma\right)$
the vorticity distribution: \begin{equation}
\gamma\left(\sigma\right)=-\frac{\mathrm{d}\bar{A}}{\mathrm{d}\sigma}\quad\mbox{with\,\,\,}\bar{A}\left(\sigma\right)=\int_{\mathcal{D}}\mathrm{d}\mathbf{r}\,\chi_{\left\{ \sigma\leq f(x)\leq f_{m}\right\} }\end{equation}
 (see section \ref{sub:Casimirs-conservation-laws} for the definition
of $\chi_{\mathcal{B}}$) %
\footnote{These definitions are different from those of section \ref{sub:Casimirs-conservation-laws}
(2D Euler equations), because with the Vlasov equation, the area of
$\mathcal{D}$ is infinite.%
}. The area $\gamma(\sigma)$ of a given $f$-level $\sigma$ (or equivalently
$\bar{A}(\sigma)$) is a dynamical invariant. The invariance of $\gamma\left(\sigma\right)$
is equivalent to the invariance of all Casimirs $C_{s}[f]$.\\

For the dynamical equilibria, there is no time dependence for $\phi$
and $\psi$. Then from (\ref{eq:Euler-Vlasov}) we conclude that any
distribution for which the distribution functions are constant over
isovalues lines of $\psi$ are dynamical equilibria. For instance,
for any $f_{0}$, distribution of the type $f(x,p)=f_{0}\left(\psi(x,p)\right)$
are dynamical equilibria (we note that this relation has to be self-consistent
as the potential defining $\phi$ is also computed from $f$).

\subsection{Equilibrium statistical mechanics of the Vlasov equation}

The equations for the equilibrium statistical mechanics of the Vlasov
equation were first written by Lynden-Bell \cite{LyndenBell:1968_MNRAS}
for self-gravitating systems; extensions and discussions of the analogy
with the 2D Euler equations for the equilibrium statistical mechanics
were first discussed in \cite{Chavanis_etal_APJ_1996}. The equilibrium
statistical mechanics can be considered following exactly the same
steps as for the 2D Euler equations, in section \ref{sec:Equilibrium-statistical-mechanics}.
Then a mean field approach will be valid and the equilibrium distribution
will be a Young measure characterized by $\rho(\sigma,x,p)$ the probability
distribution for $f$ to take the value $\sigma$ at the point $(x,p)$
of the phase space, with the normalization $\int_{-\infty}^{+\infty}\mathrm{d}\sigma\,\rho(\sigma,x,p)=1$.

Let us define the average one-particle distribution function \[
\bar{f}(x,p)\equiv\int_{-\infty}^{+\infty}\mathrm{d}\sigma\,\sigma\rho(\sigma,x,p).\]
 The mean-field variational problem defining the statistical equilibrium
is \begin{equation}
S(E,\gamma)=\sup_{\{\rho\,|\,\int_{-\infty}^{+\infty}\mathrm{d}\sigma\,\rho(\sigma,x,p)=1\}}\left\{ \mathcal{S}[\rho]\ |\ H[\bar{f}]=E,\, D[\rho](\sigma)=\int\mathrm{d}x\mathrm{d}p\,\rho(\sigma,x,p)=\gamma(\sigma)\right\} \label{eq:Vlasor_var-pb}\end{equation}
 (see \eqref{eq:MVP})\textbf{,} where $\mathcal{S}[\rho]=-\int\mathrm{d}x\mathrm{d}p\,\rho\log\rho$.
The equilibrium probability density distribution then reads: \begin{equation}
\rho(\sigma,x,p)=\frac{\mathrm{e}^{-\beta\sigma\big(\frac{p^{2}}{2}+\phi\big)-\alpha(\sigma)}}{Z_{\alpha}\big(-\beta\big(\frac{p^{2}}{2}+\phi\big)\big)},\label{eq:Vlasov_crit-pdd}\end{equation}
 where $Z_{\alpha}(u)=\int_{-\infty}^{+\infty}\mathrm{d}\sigma\ \mathrm{e}^{\sigma u-\alpha(\sigma)}$.

\subsection{Invariant measures of the Vlasov equation}

Let us now look at the evolution of the characteristic functional,
defined by \[
F[\lambda]=\left\langle \mathrm{e}^{i\int\mathrm{d}x\mathrm{d}p\lambda(x,p)f(x,p)}\right\rangle .\]
 Using a Green function formalism, in order to keep the analogy with
the 2D Euler equations, we have \[
{\bf v}[f](x,p)=\int\mathrm{d}x'\mathrm{d}p'{\mathbf{G}}(x,x';p,p')f(x',p')+(0,p),\]
 where \[
{\mathbf{G}}(x,x';p,p')=\left(0,-\frac{\mathrm{d}W}{\mathrm{d}x}(x-x')\right).\]
 By analogy with (\ref{eq:charac_funcnal}), we get \begin{align}
\frac{\partial F}{\partial t}+\int\mathrm{d}x\mathrm{d}p\ \lambda(x,p)p\frac{\partial}{\partial x}\left(\frac{\delta F}{\delta\lambda(x,p)}\right)+\notag\\
+i\int\mathrm{d}x\mathrm{d}p\mathrm{d}x'\mathrm{d}p'\ \frac{\partial}{\partial p}\left[\frac{\mathrm{d}W}{\mathrm{d}x}(x-x')\frac{\delta^{2}F}{\delta\lambda(x,p)\delta\lambda(x',p')}\right]=0\,.\notag\end{align}

The dynamics of Young measures, characterized either by $\rho\left(\sigma,x,p\right)$,
or $h\left(\lambda,x,p\right)=\log\left(\int\mathrm{d}\sigma\,\rho\left(\sigma,x,p\right)\exp\left(i\lambda\sigma\right)\right)$,
is given by \[
\frac{\partial h}{\partial t}+\mathbf{v}\cdot\mathbf{\nabla}h=0\,.\]
 The set of invariant Young measures is thus the set of Young measures
such that $h$ (or $\rho$) is constant over any isoline of the average
particle energy $\bar{\psi}$. We can define quasi-invariant Young
measures similarly to the case of the 2D Euler equations, as discussed
in section \ref{par:Quasi-invariant-Young-measures}.\\

As in the case of the 2D Euler equations, because the set of Young
measures and hence the set of invariant measures is much larger than
the set of equilibrium measures, the Vlasov equation is non-ergodic.

The discussion of the ergodicity in the framework of the Vlasov equation
has a long history, starting with the works of Lynden-Bell \cite{LyndenBell:1968_MNRAS}.
A lot of recent works have made detailed comparisons of the prediction
of the equilibrium statistical mechanics with numerical simulations
\cite{Arad_Johansson_MNRAS_2005,Arad_LB_MNRAS_2005,Antoniazzi_andco_2007PhRvE..75a1112A,AntoniazziEFR:2006_EPJB_Meca_Stat_Vlasov,Yamaguchi_2008PhRvE..78d1114Y,Levin_Pakter_Rizzati_2008PhRvE..78b1130L},
see also a detailed discussion in \cite{Chvanis_Assise_2008AIPC..970...39C}
and references therein. The qualitative results are similar to the
ones for the 2D Euler equations: whereas some cases definitely show
not so good a prediction due to the equilibrium statistical mechanics,
because of lack of ergodicity, in most cases equilibrium statistical
mechanics provides a fairly good prediction of the final self-organized
state. For instance, this theory has been used to predict the final
bunching parameter of a free-electron laser \cite{Barre_Dauxois_etal_2004_PRE_FEL}.
Moreover, the prediction skill of the equilibrium theory is expected
to be better when the dimension increases.\textbf{ }We stress that
the Vlasov equation in unbounded physical space without a confining
potential presents some specific difficulties, as then the microcanonical
measure is not defined.

\section{Perspectives\label{sec:Conclusion-and-perspectives}}

In this paper, we have studied classes of Young measures which are
invariant measures of the 2D Euler equations. These classes include
microcanonical and canonical equilibrium measures, but not only. Our
approach was to consider the problem directly from a dynamical perspective,
by looking at the evolution equations for the characteristic functional
and for the cumulant-generating functional.

Our main motivation and the interest of this approach is, first, to
study the stability of invariant measures and, second, to be able to
generalize the results to other dynamical systems, for instance the
2D Navier-Stokes equation with stochastic forces.

In sections \ref{sub:Stability-among-Young-measures} and \ref{sub:Stability-for-small-velocity},
we began the study of the stability of invariant Young measures. We
have also stressed that a more complete study of their stability requires
new results about the relaxation towards dynamical equilibria of the
2D Euler equations. Very interesting results have been obtained recently
for the relaxation towards dynamical equilibria of the Vlasov equation
\cite{Mouhot_Villani:2009}. We hope that similar results will be
obtained soon for the 2D Euler equations, and that these will be used
in the future to complete the study of the stability of both the 2D
Euler and Vlasov equations.

The class of invariant measures we describe in this paper are no-fluxes
ones. For dissipative systems, like the 2D Navier-Stokes equations
with linear friction and stochastic forces, or the 2D Euler equations
with linear frictions and stochastic forces, invariant measures will
exist which have fluxes of conserved quantities. However, for two-dimensional
flows, the energy flux is thought to converge to zero in the limit
of small forcing and dissipation, by contrast to what happens for
3D flows (anomalous dissipation). There is thus the possibility that
the limit of small forces and dissipation will be well-behaved. Some
very interesting recent mathematical results \cite{Kuksin_2004_JStatPhys_EulerianLimit}
seem to give positive insight in this direction. Our hope is that
the invariant Young measures described in this paper could be first-order
solutions in an asymptotic expansion of flux solutions for dissipative
systems. This work is a first step in this direction, by giving a
dynamical proof for sets of Young measures and studying their perturbed
dynamics.

\section*{Acknowledgments}

This work was supported through the ANR program STATFLOW (ANR-06-JCJC-0037-01)
and through the ANR program STATOCEAN (ANR-09-SYSC-014).

\bigskip{}

\section*{Appendix A: Integration in the complex plane and saddle-point approximations}

In this appendix, we evaluate the asymptotic behavior, for large $N$,
of the following integral: \begin{equation}
J_{N}\left(E,\alpha\right)=\alpha\int_{-\infty}^{+\infty}\mathrm{d}z\ \mathrm{e}^{-2iN\alpha Ez}f_{N}\left(z\right)\,\,\,\mbox{with}\,\,\, f_{N}(z)=\prod_{n=1}^{N}\left(1-i\frac{z}{\lambda_{n}}\right)^{-\frac{1}{2}},\label{eq:JN-f}\end{equation}
 where $\left\{ \lambda_{n}\right\} _{n\geq1}$ is the set of eigenvalues
of the Laplacian on the domain $\mathcal{D}$.\\

\begin{figure}
\begin{centering}
\includegraphics{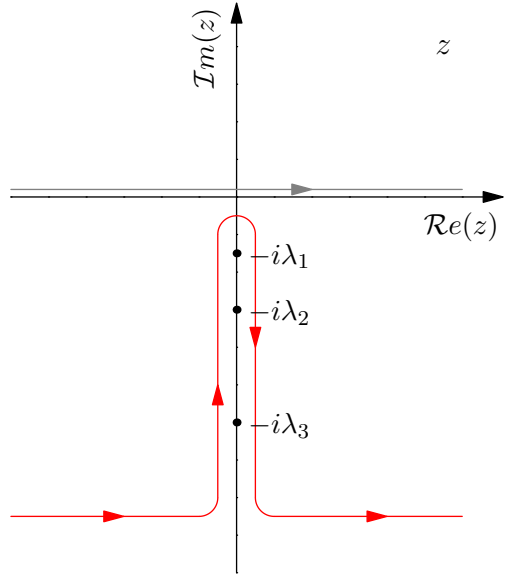} 
\par\end{centering}

\caption{$f_{N}(z)$ (\ref{eq:JN-f}) and integration contours used in the
evaluation of integral $J_{N}$ (\ref{eq:JN-f}).\label{fig:The-complex-singularities}}

\end{figure}

We first study the function $f_{N}$. It has square-root singularities
at $z=z_{n}\equiv-i\lambda_{n}$, with $1\leq n\leq N$. We consider
a complex determination of the square roots, such that $f_{N}$ has
a branch cut along the half-line $z=-ix$, with $x\geq\lambda_{1}$
(see figure \ref{fig:The-complex-singularities}). We note that each
of these singularities are integrable singularities.

We look for an asymptotic expansion (for large $N$) of $f_{N}$ .
It is a classical result \cite{Courant_Hilbert_1953_Meth_Math_Phys}
that \[
\lambda_{n}\underset{n\rightarrow\infty}{\sim}\frac{4\pi}{\left|\mathcal{D}\right|} n.\]
Then \[
\sum_{n=1}^{N}\frac{1}{\lambda_{n}}\underset{n\rightarrow\infty}{\sim}\frac{\left|\mathcal{D}\right|}{4\pi}\ln(N),\]
 and $\sum_{n=1}^{\infty}1/\lambda_{n}^{2}$ is a convergent series.

Using \[
\ln f_{N}(z)=-\frac{1}{2}\sum_{n=1}^{N}\ln\left(1-i\frac{z}{\lambda_{n}}\right)=-\frac{1}{2}\sum_{n=1}^{N}\left[-i\frac{z}{\lambda_{n}}+o\left(\frac{iz}{\lambda_{n}^{2}}\right)\right],\]
 we conclude \begin{equation}
\ln f_{N}(z)\underset{N\rightarrow\infty}{=}\frac{\left|\mathcal{D}\right|}{8\pi}(iz)\ln N+C(iz)+o\left(\frac{1}{N}\right)\quad\Rightarrow\quad f_{N}(z)\underset{N\rightarrow\infty}{\sim}N^{\frac{iz\left|\mathcal{D}\right|}{8\pi}}\tilde{C}(iz).\label{eq:fN}\end{equation}

In the vicinity of $z=-i\lambda_{1}$, a direct extension of this
result is \begin{equation}
f_{N}(-i\lambda_{1}+z)\underset{N\rightarrow\infty}{\sim}\frac{N^{\frac{\lambda_{1}\left|\mathcal{D}\right|}{8\pi}}}{(-iz)^{\frac{1}{2}}}\tilde{C}_{1}(iz),\label{eq:fN-2}\end{equation}
 where $\tilde{C}$ is analytic in the vicinity of $-i\lambda_{1}$.

Function $f_{N}\left(z\right)$ is analytical except along the branch
cut $z=-ix$, with $x\geq\lambda_{1}$. In order to compute (\ref{eq:JN-f}),
we deform the integration contour as illustrated on Figure \ref{fig:The-complex-singularities}:
the initial and deformed contours are shown in gray and red, respectively.

It is easily checked that the contribution to $J_{N}$ of the horizontal
part of the contour (see Figure \ref{fig:The-complex-singularities})
are exponentially small, for large $N$. Using the change of variable
$z=-i(x+\lambda_{1})$, we obtain \[
J_{N}(E,\alpha)\underset{N\to\infty}{\sim}\alpha\mathrm{e}^{-2N\lambda_{1}\alpha E}\int_{0}^{+\infty}\mathrm{d}x\ \mathrm{e}^{-2N\alpha Ex}\Delta f_{N}(-i (x+\lambda_{1})),\]
 where $\Delta f_{N}$ is the difference between the values of $f_{N}$
to the left and to the right of the branch cut. Using (\ref{eq:fN})
we get \[
J_{N}(E,\alpha)\underset{N\to\infty}{\sim}\alpha N^{\frac{\lambda_{1}|\mathcal{D}|}{8\pi}}\mathrm{e}^{-2N\lambda_{1}\alpha E}\int_{0}^{+\infty}\mathrm{d}x\ \mathrm{e}^{\left(-2NE+\frac{\left|\mathcal{D}\right|}{8\pi\alpha}\ln N\right)\alpha x}\Delta\tilde{C}(-i (x+\lambda_{1})).\]
 For large $N$, this last integral is clearly dominated by values
of $x$ close to zero. Then, using (\ref{eq:fN-2}):
\[J_{N}(E,\alpha)\underset{N\to\infty}{\sim}c\alpha^{1/2}N^{\frac{\lambda_{1\left|\mathcal{D}\right|}}{8\pi}}\mathrm{e}^{-2N\lambda_{1}\alpha E}\int_{0}^{+\infty}\mathrm{d}x\ \mathrm{e}^{\left(-2NE+\frac{\left|\mathcal{D}\right|}{8\pi\alpha}\ln N\right)x}x^{-\frac{1}{2}}\]
 where $c=-2i\tilde{C}_{1}(0)$. Finally
 \begin{equation}
J_{N}(E,\alpha)\underset{N\to\infty}{\sim}\frac{c\alpha^{1/2}}{\sqrt{2N}}N^{\frac{\lambda_{1\left|\mathcal{D}\right|}}{8\pi}}\frac{\mathrm{e}^{-2N\lambda_{1}\alpha E}}{\sqrt{E}}.\label{eq:JN-Asymptotics}
\end{equation}

\bigskip{}

\section*{Appendix B: Energy-enstrophy microcanonical measure}

\subsection*{B-1 Energy-enstrophy microcanonical measure from a Fourier decomposition}

In this appendix, we compute the entropy for the energy-enstrophy
ensemble, as well as related quantities. The entropy is defined by
\begin{align}
S_{K}(E,\Gamma_{2})=\lim_{N\to\infty}\left[\frac{1}{N}\log\Omega_{K,N}(E,\Gamma_{2})-C(N,\{\lambda_{n}\})\right]\,\,\,\mbox{with}\notag\\
\Omega_{K,N}\left(E,\Gamma_{2}\right)=\int\prod_{j=1}^{N}\mathrm{d}\omega_{j}\,\delta(\mathcal{E}_{N}\left[\omega\right]-E)\delta({\Gamma}_{2,N}\left[\omega\right]-{\Gamma}_{2}),\label{eq:OmegaN_Appendice}\end{align}
where $C$ does not depend on the physical parameters. It depends
only on $N$ and on the geometric factors $\{\lambda_{n}\}$ and
can be discarded as the entropy is always defined up to an arbitrary
constant.

We start by relaxing the enstrophy constraint: the Dirac delta in
enstrophy is thus replaced with a Boltzmann factor in the expression
of $\Omega_{K,N}$. Then, we compute, for $\alpha\geq0\,,$ \begin{equation}
I_{N}\left(E,\alpha\right)=\int\prod_{j=1}^{N}\mathrm{d}\omega_{j}\ \mathrm{e}^{-N\alpha\omega_{j}^{2}}\,\delta(\mathcal{E}_{N}[\omega]-E).\label{eq:IN}\end{equation}
The relation between $I_{N}\left(E,\alpha\right)$ and $\Omega_{K,N}(E,\Gamma_{2})$
shall be discussed at the end of this appendix. Let us use a representation
as an integral in the complex plane of the remaining Dirac delta function:
\begin{align}
I_{N}\left(E,\alpha\right) & =\int\prod_{j=1}^{N}\mathrm{d}\omega_{j}\ \mathrm{e}^{-N\alpha\omega_{j}^{2}}\,\delta\left(\sum_{n=1}^{N}\frac{\omega_{n}^{2}}{\lambda_{n}}-2E\right)\notag\\
 & =\frac{1}{2\pi}\int_{-\infty}^{+\infty}\mathrm{d}k_{1}\,\mathrm{e}^{-2ik_{1}E}\prod_{j=1}^{N}\int\mathrm{d}\omega_{j}\ \mathrm{e}^{-\left(N\alpha-\frac{ik_{1}}{\lambda_{j}}\right)\omega_{j}^{2}}.\notag\end{align}
 After computing the Gaussian integrals, we get \begin{align*}
I_{N}\left(E,\alpha\right) & =\frac{\pi^{\frac{N}{2}}\alpha^{-\frac{N}{2}}}{2\pi N^{\frac{N}{2}-1}}\int_{-\infty}^{+\infty}\mathrm{d}k_{1}\ \mathrm{e}^{-2iNk_{1}E}f_{N}\left(\frac{k_{1}}{\alpha}\right)\equiv\frac{\pi^{\frac{N}{2}}\alpha^{-\frac{N}{2}}}{2\pi N^{\frac{N}{2}-1}}J_{N}\left(E,\alpha\right),\end{align*}
 where $J_{N}$ and $f_{N}$ are defined by equation (\ref{eq:JN-f}),
page \pageref{eq:JN-f}.

Using result (\ref{eq:JN-Asymptotics}) of Appendix A, we obtain \begin{align}
I_{N}\left(E,\alpha\right) & \underset{N\rightarrow\infty}{\sim}C_{1}\left(N,\left\{ \lambda_{n}\right\} \right)C_{2}\left(N,\{\lambda_{n}\},\alpha\right)\frac{\exp\left[-NG\left(E,\alpha\right)\right]}{\sqrt{2E}}\,,\label{eq:large-deviation-IN}\\
 & \text{with}\quad G\left(E,\alpha\right)=2\lambda_{1}\alpha E+\frac{\ln\alpha}{2}\,,\label{eq:G_potential}\end{align}
 where $C_{1}$ depends only on $N$ and $\{\lambda_{n}\}$ (no dependence
on the physical parameters), and $C_{2}$ has no exponentially large
contributions for large $N$ ($\lim_{N\rightarrow\infty}\left(\ln C_{2}\right)/N=0$).\\

From the definition (\ref{eq:OmegaN_Appendice}) it is clear that
for small $\Delta\Gamma_{2}$ and $\Delta E$, $\Omega_{K,N}(E,\Gamma_{2})\Delta\Gamma_{2}\Delta E$
is the volume of the part of phase space with energy comprised between
$E$ and $E+\Delta E$, and enstrophy between $\Gamma_{2}$ and $\Gamma_{2}+\Delta\Gamma_{2}$.
We also note that Poincaré inequalities\textbf{ }impose $\Gamma_{2,N}[\omega]\geq2\lambda_{1}\mathcal{E}_{N}[\omega]$
and $\Gamma_{2}[\omega]\geq2\lambda_{1}\mathcal{E}[\omega]$. Then
from (\ref{eq:OmegaN_Appendice}) and (\ref{eq:IN}) we get \begin{equation}
I_{N}(E,\alpha)=\int_{2\lambda_{1}E}^{+\infty}\mathrm{d}\Gamma_{2}\,\mbox{\ensuremath{\exp}}\left(-N\alpha\Gamma_{2}\right)\Omega_{K,N}\left(E,\Gamma_{2}\right).\label{eq:IN_OmegaN}\end{equation}
It is not difficult to make for $\Omega_{K,N}(E,\Gamma_{2})$ the
same type of complex plane representation and saddle point approximation
as the one presented for $I_{N}$ in Appendix A. However, the presentation
of the computation would be tedious as it involves two complex auxiliary
variables (similar to $k_{1}$ above). We thus do not present these
computations here, but we use that a large-deviation result holds:
\begin{equation}
\Omega_{K,N}\left(E,\Gamma_{2}\right)\underset{N\rightarrow\infty}{\sim}C_{3}\left(N,\{\lambda_{n}\}\right)C_{4}\left(\{\lambda_{n}\},\Gamma_{2},N\right)\frac{\exp\left[NS_{K}\left(E,\Gamma_{2}\right)\right]}{\sqrt{2E}}+o\left(\frac{1}{N}\right).\label{eq:Volume_espace_phases_energie_enstrophie}
\end{equation}

Then using this last expression in (\ref{eq:IN_OmegaN}) and performing
a saddle point approximation, we conclude that the thermodynamic potential
\eqref{eq:G_potential} of the relaxed (canonical) ensemble is related
to the entropy $S_{K}$ through: \[
G\left(E,\alpha\right)=\min_{\Gamma_{2}\geq2\lambda_{1}E}\left\{ -S_{K}\left(E,\Gamma_{2}\right)+\alpha\Gamma_{2}\right\}.
\]
Precisely, $G(E,\cdot)$ is the Legendre-Fenchel transform of $S_{K}(E,\cdot)$.
It is a classical result that if $G$ has no singularities, then $S_{K}$
can be computed from the inverse formula $S_{K}(E,\Gamma_{2})=\min_{\alpha\geq0}\{G(E,\alpha)+\alpha\Gamma_{2}\}$
(see any textbook on convex analysis or \cite{Bouchet:2008_Physica_D}).
Using this inversion formula we get \begin{equation}
S_{K}\left(E,\Gamma_{2}\right)=\frac{1}{2}\log\left(\Gamma_{2}-2\lambda_{1}E\right)+\frac{\log2}{2}\,.\label{eq:Entropie-Energie-Enstrophie}\end{equation}
 %which is the same result as the one found from the mean-field approximation
We note that the entropy diverges for $\Gamma_{2}=2\lambda_{1}E$,
the minimal accessible enstrophy $\Gamma_{2}$ for a given energy
$E$. This could have been expected, as only the two microscopic states
$\omega=\pm\sqrt{2E}e_{1}$ verify the relation $\Gamma_{2}=2\lambda_{1}E$,
as can be readily seen from the Poincaré inequality.

\subsection*{B-2 Computation of the entropy from the Boltzmann-Gibbs entropy}

From (\ref{eq:Maxwell-Boltzmann-Entropy-Euler}) and (\ref{eq:en-en-rho-crit})
we the entropy of the equilibrium state $\rho^{*}$ of the energy-enstrophy
microcanonical measure

\begin{align}
\frac{1}{|{\mathcal{D}}|}\mathcal{S}[\rho^{\star}] & =-\sqrt{\frac{\alpha}{\pi|{\mathcal{D}}|}}\int_{\mathcal{D}}\mathrm{d}\mathbf{r}\int_{-\infty}^{+\infty}\mathrm{d}\sigma\,\mathrm{e}^{-\alpha|{\mathcal{D}}|\left(\sigma-\frac{\beta\bar{\psi}({\bf r})}{2\alpha}\right)^{2}}\left[\ln\left(\sqrt{\frac{\alpha|{\mathcal{D}}|}{\pi}}\right)-\alpha|{\mathcal{D}}|\left(\sigma-\frac{\beta\bar{\psi}({\bf r})}{2\alpha}\right)^{2}\right]\notag\\
 & =-\sqrt{\frac{\alpha}{\pi|{\mathcal{D}}|}}\int_{\mathcal{D}}\mathrm{d}\mathbf{r}\int_{-\infty}^{+\infty}\mathrm{d}\sigma\,\mathrm{e}^{-\alpha|{\mathcal{D}}|\sigma^{2}}\left[\frac{1}{2}\ln\left(\frac{\alpha|{\mathcal{D}}|}{\pi}\right)-\alpha|{\mathcal{D}}|\sigma^{2}\right]\notag\\
 & =-\frac{1}{2}\left[\ln\alpha+\ln\frac{|{\mathcal{D}}|}{\pi}\right]+\frac{1}{2}\notag\,.\end{align}
 The last two terms of the rhs being generic (the entropy being
defined up to a constant), we retain \begin{equation}
\frac{1}{\mathcal{\left|D\right|}}\mathcal{S}[\rho^{\star}]=-\frac{1}{2}\log\alpha.\label{eq:Entropy_alpha_2}\end{equation}

\bigskip{}

\section*{Appendix C: Correlation of the vorticity field}

Here, we sketch the evaluation of the order of magnitude of the two-point
correlation function, for the vorticity field obtained from the energy-enstrophy
microcanonical measure. From (\ref{eq:PDF-Omega-n}), %page \pageref{eq:PDF-Omega-n},
we see that, for the measure $\mu_{m,K}^{N}$ (\ref{eq:microcanonical_energy_enstrophy_N}),
the variance
$\langle\!\langle\omega_{n}^{2}\rangle\!\rangle_{N}=\left\langle \omega_{n}^{2}-\langle\omega_{n}\rangle_{N}^{2}\right\rangle _{N}$
of $\omega_{n}$ is of order $1/N$. Following the reasoning that
lead to equations (\ref{eq:PDF-Omega-n}), the joint probability distribution
$P_{N,n,m}(\omega_{n},\omega_{m})$ (for amplitude of modes $e_{n}$
and $e_{m}$) can be derived from \eqref{eq:Omega_N}: \begin{equation}
P_{N,n,m}(\omega_{n},\omega_{m})\underset{N\rightarrow\infty}{\sim}C\exp\left[N\log\left(\Gamma_{2}-2\lambda_{1}E-\frac{\left(\lambda_{n}-\lambda_{1}\right)}{\lambda_{1}}\omega_{n}^{2}-\frac{\left(\lambda_{m}-\lambda_{1}\right)}{\lambda_{1}}\omega_{m}^{2}\right)\right].\label{eq:PDF-jointe-omega-n}\end{equation}
 From this expression, for $n\neq m$, the correlation $\langle\!\langle\omega_{n}\omega_{m}\rangle\!\rangle_{N}=\left\langle \omega_{n}\omega_{m}-\langle\omega_{n}\rangle_{N}\langle\omega_{m}\rangle_{N}\right\rangle _{N}$
can be shown to be of order $1/N^{2}$. Now, $\omega(\mathbf{r})=\sum_{n}\omega_{n}e_{n}(\mathbf{r})$,
so $\langle\!\langle\omega(\mathbf{r})\omega(\mathbf{r'})\rangle\!\rangle_{N}$
is of order $1/N$. Therefore, for the energy-enstrophy microcanonical
measure $\mu_{m,K}$ ($N$ going to infinity), \begin{equation}
\langle\!\langle\omega(\mathbf{r})\omega(\mathbf{r'})\rangle\!\rangle=0\,.\label{eq:Correlations-Omega}\end{equation}
 We just considered the second moment of the vorticity field. However,
such a result is much more general: vorticity values at points $\mathbf{r}$
and $\mathbf{r'}$ are actually statistically independent for the
microcanonical measure, as could be easily shown by extending the
results \eqref{eq:PDF-jointe-omega-n} and \eqref{eq:Correlations-Omega}.

\bigskip{}

\section*{Appendix D: Evolution of the characteristic and cumulant-generating functionals }

In this appendix, we derive the evolution equations for the characteristic
and cumulant-generating functionals.

\subsection*{D-1 Characteristic functional}

\subsubsection*{For the 2D Euler equations}

In order to compute the evolution equation for the characteristic
functional $F[\lambda]=\langle\mathrm{e}^{i\int\lambda(\mathbf{r})\omega(\mathbf{r})\mathrm{d}\mathbf{r}}\rangle$,
we use the intermediate quantity
\[
A:=\mathrm{e}^{i\int\lambda(\mathbf{r})\omega(\mathbf{r})\mathrm{d}\mathbf{r}}.
\]
Using the Euler equation (\ref{eq:Euler_2D_Vorticity}) and an integration
by parts, we obtain \[
\frac{\partial A}{\partial t}=iA-\int\mathrm{d}{\bf r}\ \omega({\bf r})\mathbf{v}({\bf r})\cdot{\nabla}\lambda({\bf {r}).}\]
 We then use \begin{subequations}\label{eq:mean_v-omega-A} \begin{align}
\langle\mathbf{v}({\bf r})\omega({\bf r})A\rangle & =\int\mathrm{d}{\bf r'}\ \mathbf{G}\left(\mathbf{r},\mathbf{r}'\right)\langle\omega\left(\mathbf{r'}\right)\omega\left(\mathbf{r}\right)A\rangle,\\
\text{but}\quad\langle\omega({\bf r'})\omega({\bf r})A\rangle & =-\frac{\delta^{2}F}{\delta\lambda({\bf r})\delta\lambda({\bf r'})}\,,\end{align}
 \end{subequations} so that\begin{equation}
\frac{\partial F}{\partial t}+i\iint\mathrm{d}{\bf {r'}}\,\mathrm{d}{\bf {r}}\ {\nabla}\lambda({\bf r})\cdot{\bf G}({\bf r},{\bf r'})\frac{\delta^{2}F}{\delta\lambda({\bf r})\delta\lambda({\bf r'})}=0.\label{eq:Characteristic-Functional-Appendice-2}
\end{equation}

\subsubsection*{For the perturbation of an equilibrium of the 2D Euler equations }

We now apply the same tools to the case of the 2D Euler equations
near a dynamical equilibrium. Consider $\Omega$ a solution to the
2D Euler equations, with $\Omega=\Omega_{0}+\varepsilon\omega\,,\,\mathbf{V}=\mathbf{V_{0}}+\varepsilon\mathbf{v}$
and $\mathbf{V_{0}}\cdot\nabla\Omega_{0}=0$ (see section \ref{sub:Stability-for-small-velocity}).

We want to determine the evolution equation for the characteristic
functional $F[\lambda]=\langle\mathrm{e}^{i\int\lambda(\mathbf{r})\omega(\mathbf{r})\mathrm{d}\mathbf{r}}\rangle$,
just as we did in the case of the 2D Euler equations. Defining $A:=\mathrm{e}^{i\int\lambda(\mathbf{r})\omega(\mathbf{r})\mathrm{d}\mathbf{r}}$
and using \eqref{eq:linearized-transport}, page \pageref{eq:linearized-transport},
we get \begin{align}
\frac{\mathrm{d}A}{\mathrm{d}t} & =-i\int\lambda\left({\bf v}\cdot\nabla\Omega_{0}+{\bf V_{0}}\cdot\nabla\omega+\varepsilon{\bf v}\cdot\nabla\omega\right)A\nonumber \\
 & =i\int\nabla\lambda\cdot\left({\bf v}\Omega_{0}+{\bf V_{0}}\omega+\varepsilon{\bf v}\omega\right)A\,.\notag\end{align}
 Since \begin{equation}
\langle\mathbf{v}({\bf r})A\rangle=\int\mathrm{d}{\bf r'}\ \mathbf{G}\left(\mathbf{r},\mathbf{r}'\right)\langle\omega\left(\mathbf{r'}\right)A\rangle=-i\int\mathrm{d}{\bf r'}\ \mathbf{G}\left(\mathbf{r},\mathbf{r}'\right)\frac{\delta F'}{\delta\lambda({\bf r})},\label{eq:mean_omega-A}\end{equation}
 we conclude

\begin{equation}
\frac{\partial F}{\partial t}+\iint{\nabla}\lambda({\bf r})\cdot\left[{\bf G}({\bf r},{\bf r'})\left(\varepsilon i\frac{\delta^{2}F}{\delta\lambda({\bf r})\delta\lambda({\bf r'})}-\Omega_{0}({\bf r})\frac{\delta F}{\delta\lambda({\bf r'})}\right)-{\bf V_{0}}({\bf r})\frac{\delta F}{\delta\lambda({\bf r})}\right]\ \mathrm{d}{\bf {r'}}\,\mathrm{d}{\bf {r}}=0.\label{eq:Characteristic-functional-closeto equilibrium}\end{equation}

\subsection*{D-2 Cumulant-generating functional}

By definition, the cumulant-generating functional is $H=\ln F$. Therefore\[
\frac{\partial H}{\partial t}=\frac{1}{F}\frac{\partial F}{\partial t}\:,\quad\frac{\delta H}{\delta\lambda({\bf r})}=\frac{1}{F}\frac{\delta F}{\delta\lambda({\bf r})}\:,\,\,\,\mbox{and}\,\,\,\frac{\delta^{2}H}{\delta\lambda({\bf r})\delta\lambda({\bf r'})}=\frac{1}{F}\frac{\delta^{2}F}{\delta\lambda({\bf r})\delta\lambda({\bf r'})}-\frac{1}{F^{2}}\frac{\delta F}{\delta\lambda({\bf r})}\frac{\delta F}{\delta\lambda({\bf r'})}.\]

Hence, using (\pageref{eq:Characteristic-Functional-Appendice-2}),
\[
\frac{\partial H}{\partial t}+i\iint{\nabla}\lambda({\bf r})\cdot{\bf G}({\bf r},{\bf r'})\left(\frac{\delta^{2}H}{\delta\lambda({\bf r})\delta\lambda({\bf r'})}+\frac{\delta H}{\delta\lambda({\bf r})}\frac{\delta H}{\delta\lambda({\bf r'})}\right)\,\mathrm{d}{\bf {r'}}\,\mathrm{d}{\bf {r}}=0.\]

For a small perturbation of a dynamical equilibrium, using (\pageref{eq:Characteristic-functional-closeto equilibrium}),
we get

\[
\frac{\partial H}{\partial t}+\iint{\nabla}\lambda({\bf r})\cdot\left[{\bf G}({\bf r},{\bf r'})\left(\varepsilon i\left(\frac{\delta^{2}H}{\delta\lambda({\bf r})\delta\lambda({\bf r'})}+\frac{\delta H}{\delta\lambda({\bf r})}\frac{\delta H}{\delta\lambda({\bf r'})}\right)-\Omega_{0}({\bf r})\frac{\delta H}{\delta\lambda({\bf r'})}\right)-{\bf V_{0}}({\bf r})\frac{\delta H}{\delta\lambda({\bf r})}\right]\mathrm{d}{\bf {r'}}\,\mathrm{d}{\bf {r}}=0.\]

\bigskip{}

\section*{Appendix E: Uniqueness of the Vlasov statistical equilibria for repulsive
convex potentials}

In this appendix, we prove that, for a repulsive convex potential $W$,
the microcanonical variational problem for the Vlasov equation has
a unique solution. This has the following consequences: for a repulsive
potential, no phase transition exists and no ensemble inequivalence
exists. The argument is extremely simple; it is based on the concavity
of $\mathcal{G}[\rho] = \mathcal{S}[\rho]-\beta H-\int\mathrm{d}\sigma\mathrm{d}x\mathrm{d}p\,\alpha(\sigma)\rho(\sigma,x,p)$.
The concavity of $\mathcal{G}$ implies the uniqueness of the critical
point of $\mathcal{G}$ (the uniqueness of the equilibrium in the
grand canonical ensemble where $\beta$ and $\alpha$ are the control
parameters) and thus the absence of phase transitions in the grand-canonical
ensemble. Then a classical result of convex analysis \cite{Rockafellar_1970_Convex_analysis}
(see also a simple discussion in \cite{Bouchet:2008_Physica_D}) implies
that there is a one-to-one relation between the constraints $\left(E,\gamma\left(\sigma\right)\right)$
and the Lagrange multipliers $\left(\beta,\alpha\left(\sigma\right)\right)$.
Then for any energy $E$ and distribution $\alpha\left(\sigma\right)$,
the microcanonical variational problem has a unique solution and there
is no phase transition either in the microcanonical ensemble.
\par
As $\mathcal{S}$ is strictly concave and $\int\mathrm{d}\sigma\mathrm{d}x\mathrm{d}p\,\gamma(\sigma)\rho(\sigma,x,p)$
is linear, in order to prove the strict concavity of $\mathcal{G}$
it is sufficient to prove that $-\beta H$ is concave.

It is well-known that for systems with kinetic energy $E_{c}=p^{2}/2$,
the inverse temperature $\beta$ is positive. That this is necessary
can be seen directly from the normalizability of the equation for
the critical state \eqref{eq:Vlasov_crit-pdd}. Systems with possible
negative temperature states are the ones with bounded phase space
(see for instance the case of the point vortex model \cite{Onsager:1949_Meca_Stat_Points_Vortex}).
Then, in order to prove the concavity of $-\beta H$ it is sufficient
to prove the convexity of $H$ \eqref{eq:Energie_Continue}, or equivalently
to prove the convexity of the potential \[
V[f]=\frac{1}{2}\int\mathrm{d}x\mathrm{d}p\, f\phi[f]=\frac{1}{2}\int\mathrm{d}x\mathrm{d}x'\, m(x)m(x')W(x-x'),\]
 with $m(x)=\int\mathrm{d}p\, f(x,p)$. We also remark that if the
system were confined by\textbf{ }some external potential,\textbf{
}because this would appear as a linear term in the functional, the
convexity of $H$ would not be affected.

The second-order variations of the potential read \[
\delta^{2}V=\frac{1}{2}\int\mathrm{d}x\mathrm{d}x'\,\delta m(x)\delta m(x')W(x-x').\]
 Resorting to Fourier transforms, $\delta m(x)=\int\mathrm{d}k\,\delta m_{k}\mathrm{e}^{ikx}$,
we see that convexity of $V$ is equivalent to positivity of $W_{k}$,
the Fourier transform of W: \[
W_{k}=\int_{0}^{+\infty}\mathrm{d}x\, W(x)\cos(kx);\]
 recalling that $W$ is even.

We have $W_{k}=\sum_{n=0}^{\infty}W_{k,n},$ where
\begin{align}
W_{k,n} & = \int_{\frac{2n\pi}{k}}^{\frac{2(n+1)\pi}{k}}\mathrm{d}x\, W(x)\cos(kx) \notag\\
 & = \frac{1}{k}\int_{0}^{\frac{\pi}{2}}\mathrm{d}x\,\cos(x)\left[W\left(\frac{x+2n\pi}{k}\right)-W\left(\frac{\pi-x+2n\pi}{k}\right)-W\left(\frac{\pi+x+2n\pi}{k}\right)+W\left(\frac{2\pi-x+2n\pi}{k}\right)\right] \notag\\
 & = \frac{1}{k}\int_{0}^{\frac{\pi}{2}}\mathrm{d}x\,\cos(x)\int_{\frac{x+2n\pi}{k}}^{\frac{\pi-x+2n\pi}{k}}dy\,\left[W'(y+\pi)-W'\left(y\right)\right]. \notag
 \end{align}

Now, using that $W$ is convex, $W'$ is increasing and thus $W_{k,n}\geq0$.
Hence $\forall\, k\,\,\, W_{k}\geq0$ and the potential energy functional
is convex. Therefore, $H$ is convex, and the uniqueness of the statistical
equilibria is proved.

\smallskip{}
The same type of arguments could be derived in dimensions $d$ larger
than 1. We treat, for example, the case $d=3$. Assuming the interaction
potential to be isotropic, $W=W(r)$, we have \[
W(\mathbf{k})\equiv\frac{1}{(2\pi)^{3}}\int\mathrm{d}\mathbf{r}\, W(\mathbf{r})\mathrm{e}^{i\mathbf{k}\cdot\mathbf{r}}=\frac{1}{(2\pi)^{2}k}\int_{0}^{+\infty} \mathrm{d}r\, rW(r)\sin(kr).
\]
Clearly, we can use the same reasoning as before, applying it to
$rW(r)$ rather than to $W(x)$. We thus conclude that, in dimension
3, if $W$ is a repulsive isotropic potential and if $rW(r)$ is convex,
then $\forall\;\mathbf{k}\,\,\, W(\mathbf{k})\geq0$, the potential
energy functional is convex and hence, for any value of the constraints,
the statistical equilibria are unique and no phase transition exists.

Let us discuss the special case of algebraic potentials $W(r)=C/r^{\alpha}$
in dimension $3$. We are interested in systems with long-range interactions
(non-integrable potentials); then we suppose $\alpha\leq2$. The condition
$rW\left(r\right)$ convex is then $\alpha\geq1$ (for $\alpha<1$,
the Fourier transform of the potential would not be defined). Then
the preceding argument applies to values of $\alpha$ with $1\leq\alpha\leq2$,
including for instance Coulomb potentials.

\bibliographystyle{amsplain}
\bibliography{FBouchet,Long_Range,Meca_Stat_Euler,Ocean,Experimental_2D_Flows,Euler_Stability,Jupiter,Turbulence_2D,Quasilinear,Cascade,Euler2D-Linearized,rings,Statistical-Mechanics,NS-Stochastic,Kinetic-Theories-Turbulence,Maths}

\end{document}